\documentclass[a4paper]{aa}

\usepackage{graphicx,natbib}
\usepackage{txfonts}
\usepackage{subfigure}
\usepackage{dblfloatfix}
\usepackage{fixltx2e}
\usepackage[usenames,dvipsnames]{xcolor}
\usepackage{aalongtable}
\usepackage{epstopdf} 
\usepackage{lscape}

\newcommand{\bz}{$\langle B_\mathrm{z} \rangle$}

\newcommand{\kms}{km\,s$^{-1}$}

\newcommand*{\rom}[1]{\expandafter\@slowromancap\romannumeral #1@}

\begin{document}

\title{Magnetic fields of young solar twins
\thanks{Based on observations made with the HARPSpol instrument on
the ESO 3.6 m telescope at La Silla (Chile), under the program ID 091.D-0836. Also based on observations obtained at the Bernard Lyot Telescope (TBL, Pic du Midi, France) of the Midi-Pyr\'{e}n\'{e}es Observatory, which is operated by the Institut National des Sciences de l'Univers of the Centre National de la Recherche Scientifique of France.}}

\author{L.~Ros\'{e}n  \inst{1}
   \and O.~Kochukhov \inst{1}
   \and T.~Hackman \inst{2}
   \and J.~Lehtinen \inst{2,3}
  }
\institute{Department Physics and Astronomy, Uppsala University, Box 516, 751 20 Uppsala, Sweden
              \and Department of Physics, P.O. Box 64, FI-00014 University of Helsinki
              \and ReSoLVe Centre of Excellence, Aalto University, Department of Computer Science, PO Box 15400, 00076 Aalto, Finland
}

\date{Received 7 March 2016 / Accepted 10 May 2016}

\titlerunning{Magnetic fields of young Suns}

\abstract
{}
{The goal of this work is to study the magnetic fields of six young solar-analogue stars both individually, and collectively, to search for possible magnetic field trends with age. If such trends are found, they can be used to understand magnetism in the context of stellar evolution of solar-like stars and, the past of the Sun and the solar system. This is also important for the atmospheric evolution of the inner planets, Earth in particular.}
{We used Stokes $IV$ data from two different spectropolarimeters, NARVAL and HARPSpol. The least-squares deconvolution multi-line technique was used to increase the signal-to-noise ratio of the data. We then applied a modern Zeeman-Doppler imaging code in order to reconstruct the magnetic topology of all stars and the brightness distribution of one of our studied stars. }
{Our results show a significant decrease in the magnetic field strength and energy as the stellar age increases from 100~Myr to 250~Myr while there is no significant age dependence of the mean magnetic field strength for stars with ages 250-650 Myr. The spread in the mean field strength between different stars is comparable to the scatter between different observations of  individual stars. The meridional field component has the weakest strength compared to the radial and azimuthal field components in 15 out of the 16 magnetic maps. It turns out that 89-97\% of the magnetic field energy is contained in $l$=1-3. There is also no clear trend with age and distribution of field energy into poloidal/toroidal and axisymmetric/non-axisymmetric components within the sample. The two oldest stars in this study do show a twice as strong octupole component compared to the quadrupole component. This is only seen in one out of 13 maps of the younger stars. One star, $\chi^1$~Ori displays two field polarity switches during almost 5 years of observations suggesting a magnetic cycle length of either 2, 6 or 8 years.} 
{}


\maketitle

\section{Introduction}
\label{intro}

Magnetic fields play a key role in stellar evolution. Since the protostellar gas clouds in which stars are born often have magnetic field lines embedded into them, the star can acquire a seed magnetic field from the very beginning. As the star is formed by a contraction of the gas, the magnetic field lines in the gas will also be compressed and therefore strengthened. In order to conserve angular momentum the contraction of the gas will lead to an increase in the angular velocity of the protostar. Without a magnetic field, the protostar would continue to increase its angular velocity unhindered as the gas continues to contract and would eventually spin faster than the surrounding gas. If, however, there are magnetic field lines connecting the protostar and the disk, the protostar will not be able to rotate significantly faster than the disk since that would cause a bending of the field lines. The magnetic stress in the field lines prevents them from bending and twisting and the protostar slows down through a transfer of angular momentum from the star to the disk (e.g. \citealt{Donati2009} and references therein). 

The angular momentum loss will continue throughout the stars life because of the magnetised stellar wind \citep[e.g.][]{Matt2012,Reville2015}. Again, to prevent the field lines from bending, angular momentum is transferred from the star and outwards \citep{Schatzman1962} to a radius where the magnetic energy is less than the kinetic energy of the stellar wind. 

Apart from the effect on rotation, a magnetic field can also cause an increase in emission of high energy radiation and bursts of mass ejections resulting in a stronger stellar wind. This will not only affect the star, but also its surroundings. 

In contrast to hot stars, which are believed to have fossil and therefore more or less constant fields, the magnetic fields of cool stars are constantly changing and evolving because cool stars are actively generating a magnetic field in their interiors. For example, the Sun's magnetic field is not constant, neither on a global scale nor on a local scale. Small-scale magnetic spots are continuously appearing and disappearing on the Sun's surface with a typical lifetime of days to months and with magnetic field strengths of a few kG. The number of sunspots and their positions are also not constant but changing periodically. The spot number varies with an 11~year cycle at the same time as the spot activity is migrating towards the equator on each hemisphere. This is also connected to the behaviour of the global field since it switches polarity in the middle of each cycle giving it a full length of 22 years. 

The solar magnetic field generation is not completely understood, but there are theories and numerical models which are able to form magnetic bipolar regions \citep{Warnecke2013}, sustain realistic sunspots \citep{Rempel2009} and reproduce, with varying degree of success, certain global features of the solar field (e.g. \citealt{Brun2015} and references therein). It is believed that one of the main driving mechanisms for the global field of a cool star is a dynamo mechanism \citep[e.g.][]{Charbonneau2010}, which is essentially an interplay between differential rotation and convection where the magnetic field strength depends on the Rossby number, which in turn depends on the rotation period. The Sun today has a relatively long rotation period of about 25 days giving it a global field strength of only a few G. Since the magnetic field slows down the stellar rotation, the Sun should have rotated much faster when it was young. This, in turn, implies that the global field strength and the magnetic activity should have been higher. Previous observational studies have shown that the activity level of cool stars indeed decreases with increasing age \citep[e.g.][]{Vidotto2014}. 

A higher solar activity could have had a large impact on the young solar system. High energy radiation might interact with molecules in the upper parts of a planetary atmosphere causing changes in the chemical composition resulting in different heating and cooling effects of the planet \citep[e.g.][]{Kulikov2007,Lammer2012}. Perhaps this could have happened to the Earth's atmosphere and it could also have removed most of Mars' atmosphere (\citealt{Wood2006} and references therein). The position of the habitable zone is not only determined by the temperature of the star, but also by its magnetic field and many other parameters \citep[e.g.][]{Cockell2016}. 

The activity of a star can be investigated via several different proxy indicators, for example X-ray emission, photometric variability, and emission in the line cores of chromospheric lines. A magnetic field can also be directly studied by detecting magnetically induced polarisation in spectral lines affected by the Zeeman effect. A time-series of such polarisation profiles can be used to reconstruct the magnetic field topology by applying the Zeeman-Doppler imaging technique \citep[ZDI,][]{Brown91}. Circular polarisation is stronger than linear polarisation, up to 10 times, and linear polarisation has only been detected in cool stars a few times previously \citep{Kochukhov11,Rosen13} and only one cool star has been modeled using all four Stokes parameters \citep{Rosen2015}. The solar-like stars investigated in this study do not have strong enough magnetic fields for linear polarisation to be detected at the S/N limit of the current instrumentation and telescopes.

In this study, we investigate six solar analogue stars with an age of about 100--650 Myr using ZDI. These objects are all part of the ``Sun in Time'' sample \citep{Gudel2007} which is a sample of carefully selected stars with masses similar to the Sun, but with ages varying from 0.1~Gyr to 8.5~Gyr. This sample allows to probe magnetic field and related stellar activity in the context of solar evolution, largely separating the trends with age/rotation from the dependence on stellar mass. We have derived magnetic maps of all six stars from 16 sets of observations in total. Each star has been studied individually, but also together as a sample in order to investigate the overall characteristics of magnetic fields in young solar analogues as a function of age.

\section{Observations}

We retrieved archival observational data using PolarBase \citep{Petit2014} of five solar-like stars; EK~Dra, HN~Peg, $\pi^1$~UMa, $\chi^1$~Ori and $\kappa^1$~Cet. PolarBase contains observational data from the spectropolarimeter NARVAL at the 2~m Telescope \textit{Bernard Lyot} at Pic du Midi Observatory and from the spectropolarimeter ESPaDOnS at the 3.6~m Canada-France-Hawaii Telescope. We have used observational data from NARVAL in this study. We extracted two sets of observations of EK~Dra, from 2007 and 2012, five sets of observations of HN~Peg, one for each year from 2007--2011, one set of $\pi^1$~UMa from 2007, four sets of $\chi^1$~Ori from 2007, 2008, 2010 and 2011 and one set of $\kappa^1$~Cet from 2012. 

We also initiated an observational program called ``Active Suns'' where we observed a small sample of stars with the spectropolarimeter HARPSpol at the ESO 3.6-m telescope in La Silla, Chile during September 2013. In this study, we use data for HN~Peg, BE~Cet and $\kappa^1$~Cet obtained in the context of that programme. Another set of active stars from that programme was investigated in the study by \citet{Hackman2015}.  
Fundamental information about the targets and observations can be found in Table~\ref{tab_obs}.

NARVAL is an echelle spectrograph with a polarimetric mode. Three Fresnel rhombs are used to analyse the different polarised states \citep{Petit2008}. It is mounted at the Cassegrain focus in order to reduce the number of reflections and avoid additional spurious polarisation before the light enters the spectrograph. The spectrograph covers the entire optical spectrum from 3700--10500 \AA \ and has a resolving power of about 65\,000. The data were automatically reduced with the Libre-ESpRIT software \citep{Donati97}. In addition, we used our own IDL routines to do continuum normalisation by fitting a global smooth function to the Stokes $I$ spectra. 

HARPSpol is also mounted at the Cassegrain focus but uses achromatic wave plates instead of Fresnel rhombs. It consists of two polarimeters, one for circular polarisation and one for linear polarisation. In this study only the circular polarisation mode was used. The wavelength coverage is not as large as for NARVAL, 3600--6910 \AA, with an 80~\AA \ gap at 5290~\AA, but HARPSpol offers a higher resolving power of about 110\,000 \citep{Snik2011,Piskunov2011}. The data were reduced using the REDUCE package \citep{Piskunov2002}. The continuum normalisation was then done using a similar set of IDL routines as for the NARVAL spectra.

\begin{table*}
\caption{Information about the stars in this study. Most parameters are taken from previous studies while some are derived in this study. A detailed description of the stellar ages can be found in section~\ref{age}.}
\label{tab1}
\centering
\resizebox{\textwidth}{!}{
\begin{tabular}{ccccccccccccc}
\hline\hline
Star     & Star               &  $T_{\rm{eff}}$    &  $v_{\rm e}\sin i$ & Mass & Radius &Incl. & $P_{\rm{rot}}$      &  $\log L_{\rm X}$$^{4}$ & $\log R'_{\rm HK}$ & Age range & Membership &Age  \\ 
name    & HD             & $(K)$                  &   (\kms)     & $(M_{\odot})$ & $(R_{\odot})$         & (deg.) & $(d)$   & (erg/s)   & min/max  & (Myr) &  & (Myr)         	\\			 
\hline
EK~Dra & 192333 & 5845$^1$	         & 16.8$^1$   & 1.044$^{2}$	& 0.97$^{2}$ & 60$^3$ &  2.6$^3$ & 29.93        &   -4.23/-4.15      & 100-125  & Pleiades $^{5}$  &100 \\
	 
HN~Peg  & 206860  & 5974$^1$          &  10.6$^1$  & 1.103$^{2}$ & 1.04$^{2}$ & 75$^{6}$ & 4.6$^{6}$  & 29.12  & -4.44/-4.42  & 200-400  &  Hercules-Lyra $^{7}$  & 250     \\

$\pi^1$~UMa  & 72905  &  5873$^{8}$        &   11.2$^{9}$  & 1.00$^{8}$ & 0.96$^{10}$ & 60 & 4.9$^{11}$  & 29.10   & -4.40/-4.33 & 200-600  &  Ursa major $^{5}$  & 300       \\
				  
$\chi^1$~Ori  & 39587  & 5882$^1$          &    9.8$^1$   & 1.028$^{2}$   & 1.05$^{2}$ & 65 & 5.08$^{10}$ & 28.99        & -4.43/-4.42  & 300-600  & Ursa major $^{12}$    &300     \\
				  
BE~Cet    &1835       & 5837$^1$           &     7.0$^1$   & 1.062$^{2}$     & 1.00$^{2}$  & 65 & 7.65$^{10}$ & 29.13 & -4.44/-4.43 & 400-625  &  Hyades $^{13}$    & 600   \\
				  
$\kappa^1$~Cet & 20630    & 5742$^1$    &    5.2$^1$   & 1.034$^{2}$  & 0.95$^{2}$ & 60 & 9.2$^{11}$ & 28.79 & -4.47/-4.42  & 300-750   &   -  & 650   \\
				
\hline
\end{tabular}
}
\tablebib{
(1)~\citet{Valenti2005};~(2)~\citet{Takeda2007};~(3)~\citet{Strassmeier1998};~(4)~\citet{Gudel1998}~(5)~\citet{Montes2001};~(6)~\citet{Boro-Saikia2015};~(7)~\citet{Eisenbeiss2013};~(8)~\citet{Gonzalez2010};~(9)~\citet{Martinez-Arnaiz2010};~(10)~\citet{Gudel2007};~(11)~\citet{Messina2003};~(12)~\citet{King2003};~(13)~\citet{Montes01}.
}
\end{table*}

\section{The stellar sample}
\label{sample}
All the stars are chosen from the so called ``Sun in Time'' sample \citep{Gudel2007}. This sample contains stars which have parameters similar to the Sun, but with ages ranging from 0.1 to 8.5 Gyr. The goal of the ``Sun in Time'' study is to investigate the Sun's past and future in order to understand the evolution of the Sun and solar system. All the stars in our study are younger than the Sun with ages spanning from about 100 to 650 Myr. 

In Table~\ref{tab1} a summary of the stellar parameters is given. The listed values of effective temperature, projected rotational velocity, mass and radius are all taken from the literature. For the inclination, we adopted literature values for EK~Dra and HN~Peg, while the inclinations for the other four stars were derived in this study. To determine the inclination we performed magnetic inversions for a grid from 0 to 90$^\circ$ with a step of $5^\circ$ and selected the value resulting in the smallest deviation between the model profiles and observational profiles. It should be noted that for most observational sets, several inclinations resulted in similar deviations. A similar procedure was used to determine the rotation period, but instead of having a full grid, we tested previously reported rotation periods. The adopted values and their original source are listed in Table~\ref{tab1}.  

All these stars were included in the Mount Wilson survey between 1966 and 1991 where the Ca\,{\sc ii} H\&K line core emission was measured. \citet{Baliunas1995} used these observations to calculate the ratio between the core emission and the continuum, the $S$-index, in order to investigate activity periods. They found all stars to be active, but no cycle periods could be derived for EK~Dra, $\pi^1$~UMa and $\chi^1$~Ori. Activity cycle periods of $9.1 \pm 0.3$ years and $5.6 \pm 0.1$ years were found for BE~Cet and $\kappa^1$~Cet respectively. HN~Peg was suggested to have a cycle of $6.2 \pm 0.2$ years, but with a high False Alarm Probability. In a later study \citet{Choi2015} determine a cycle period of $12.60\pm 0.52$ years using the same data.

Temperature maps of EK~Dra have previously been derived through a Doppler imaging study \citep{Strassmeier1998} and \citet{Jarvinen2005} produced a map showing the spottedness of the star by using photometric light curves. The magnetic field of HN~Peg was recently reconstructed in a study by \citet{Boro-Saikia2015}. All stars except BE~Cet were also included in the study by \citet{Vidotto2014} which looked for magnetic activity trends using a larger sample of F, G, K and M-dwarfs with ages from 1~Myr to 10~Gyr. Four stars (HN~Peg, $\pi^1$~UMa, $\chi^1$~Ori, $\kappa^1$~Cet) were also part of the BCool spectropolarimetric survey \citep{Marsden2014}.

\subsection{Age estimates}
\label{age}
Since one of the focuses of this study is to investigate how the stellar activity of a Sun-like star evolves with time, it is important to have the correct age estimates of these stars, and to order them relative to each other. Determination of stellar ages is a complex problem and several different age indicators and models can be used.

Activity indicators can also provide information about the age of a star since activity and age are believed to be correlated. One such quantity is X-ray emission. It has been shown that the X-ray luminosity, $L_{\rm X}$, correlates with the large scale magnetic field \citep{Vidotto2014}. Included in Table~\ref{tab1} is therefore $\log L_{\rm X}$ of all these stars from the study by \citet{Gudel1998}. The $\log L_{\rm X}$ is decreasing with increasing rotation period for all stars except BE~Cet. \citet{Hempelmann1995} derives $\log L_{\rm X}$ for all stars except EK~Dra. $\pi^1$~UMa and $\chi^1$~Ori have the highest $\log L_{\rm X}$=29.10~erg/s while HN~Peg has a slightly lower value of 29.00~erg/s and BE~Cet and $\kappa^1$~Cet an even lower value of 28.90~erg/s.

Another activity proxy is the emission in the Ca\,{\sc ii} H\&K line core. This can be quantified as the chromospheric activity indicator $\log R'_{\rm HK}$ which is calculated from the $(B-V)$ colour and $S$-index. In Table~\ref{tab1} we list the min/max values of each star as a result from a literature search. As mentioned in section~\ref{sample}, all stars in our study were part of the Mount Wilson survey. We used the $(B-V)$ colours and $S$-index from \citet{Baliunas1995} to calculate $R'_{\rm HK}$, following the methodology described in \citet{Noyes1984} to remove the photospheric contribution. Values for all stars can also be found in \citet{Rocha-Pinto2004}. Parts of this sample is also included in studies by \citet{Wright2004}, \citet{Mamajek2008} and \citet{Vican2012}. From the ranges listen in Table~\ref{tab1}, it can be seen that EK~Dra generally has a higher $\log R'_{\rm HK}$ compared to the other stars, which all have very similar values. \citet{Song2004} argues that ages derived form this index often are overestimated, compared to values from Li measurements and X-ray luminosity, at least for stars with ages younger than and similar to the Pleiades but perhaps also for older stars.  
 
Both $\log L_{\rm X}$ and $\log R'_{\rm HK}$ varies for the same star during different observation epochs since the activity is not constant. Especially for $\log R'_{\rm HK}$, many of the stars, except EK~Dra, have similar or overlapping values. Therefore, the estimated ages using these proxies have a large spread and are also overlapping, as can be seen in Table~\ref{tab1}. Hence, these proxies are probably not too reliable as age indicators for a small sample of stars with ages of a few 100~Myr. All stars except $\kappa^1$~Cet are believed to be members of stellar kinematic groups, listed in Table~\ref{tab1}. Membership is not based on activity, and our adopted age estimates are therefore heavily weighted by the average age of the group.

\subsubsection{EK~Dra}
 
EK~Dra is believed to be a member of the Pleiades moving group \citep{Montes2001} which has an average age of 100-125 Myr \citep{Meynet1993,Stauffer1998}. \citet{Gudel2007} assigns it with an age of 100~Myr. We have also adopted this age which makes EK~Dra the youngest star in our study.

\subsubsection{HN~Peg}

The literature values of the age of HN~Peg range from about 200~Myr up to 400~Myr, depending on the used method. HN~Peg is believed to be a member of the Hercules-Lyra association which has an average gyrochronological age of $257 \pm 46$~Myr \citep{Eisenbeiss2013}. In a study by \citet{Vican2012}, the age of HN~Peg is determined by X-ray emission, $\log (L_{\rm X}/L_{\rm bol})$, and a $\log R'_{\rm HK}=-4.440$ to 300 and 400~Myr respectively. \citet{Zuckerman2009} argue that the age of HN~Peg should be about 200~Myr since the logarithmic ratio of X-ray luminosity to bolometric luminosity is similar to the value of the Pleiades. In \citet{Gudel2007} the adopted age is 300~Myr based on a rotation period of 4.86~d. 

We have adopted an age of 250~Myr, which is close to the estimated average age of the Hercules-Lyra association.

\subsubsection{$\pi^1$~UMa}

Many studies suggest $\pi^1$~UMa to be a member of the Ursa Major moving group \citep{Montes2001,Montes01,King2003}. The normally quoted age of the moving group is 300~Myr \citep{Soderblom1993}, but \citet{King2003} suggests an age of $500 \pm 100$~Myr. \citet{Vican2012} also investigated this star and determined the age to be 300~Myr using $\log R'_{\rm HK}=-4.400$. \citet{Mamajek2008} derived an age of about 200~Myr from $\log R'_{\rm HK}=-4.375$. The adopted age for $\pi^1$~UMa is 300~Myr which is the average age of the Ursa Major moving group.

\subsubsection{$\chi^1$~Ori}

$\chi^1$~Ori is also believed to be a member of the Ursa Major moving group \citep{King2003}. \citet{Mamajek2008} derived an age of about 400~Myr from $\log R'_{\rm HK}=-4.426$. We have used the same age as the group, 300~Myr \citep{Gudel2007}.

\subsubsection{BE~Cet}

BE~Cet is believed to be a member of the Hyades cluster \citep{Montes01} which has a cluster age of about 625~Myr \citep{Perryman1998}. In \citet{Gudel2007} it is therefore given an age of 600~Myr. \citet{Vican2012} estimates a slightly lower age of 400~Myr using both $\log R'_{\rm HK}=-4.440$ and $\log (L_{\rm X}/L_{\rm bol})=-4.58$. We used an age of 600~Myr for BE~Cet.

\subsubsection{$\kappa^1$~Cet}
$\kappa^1$~Cet is not known to be a member of any cluster and in \citet{Gudel2007} it is assumed to have an age of 750~Myr based on an age-rotation relationship adopting the same rotation period as we have used in this study. \citet{Lachaume1999} and \citet{Mamajek2008} derived a lower age of about 300~Myr from $\log R'_{\rm HK}=-4.420$. In another study \citet{Ribas2010} calculate the mean H$\alpha$ activity of $\kappa^1$~Cet and find it comparable to stars in the Hyades \citep{Lyra2005}. The measured X-ray luminosity by \citet{Gudel1997} also agrees well with members of the Hyades and Praesepe \citep{Barrado1998} which has a similar age as the Hyades. \citet{Vican2012} also investigated this star and determined the age to be 450 and 500~Myr using $\log (L_{\rm X}/L_{\rm bol})=-4.58$ and $\log R'_{\rm HK}=-4.470$ respectively. 

Since $\kappa^1$~Cet is not known to be a member of any moving group, the age was estimated from other indicators. Many parameters of $\kappa^1$~Cet are comparable to stars in the Hyades. A direct comparison with BE~Cet would suggest a slightly older age and we have therefore adopted an age of 650~Myr for this star, as this is close to the average age of the Hyades.

\begin{figure*}
\centering
\includegraphics[scale=0.5]{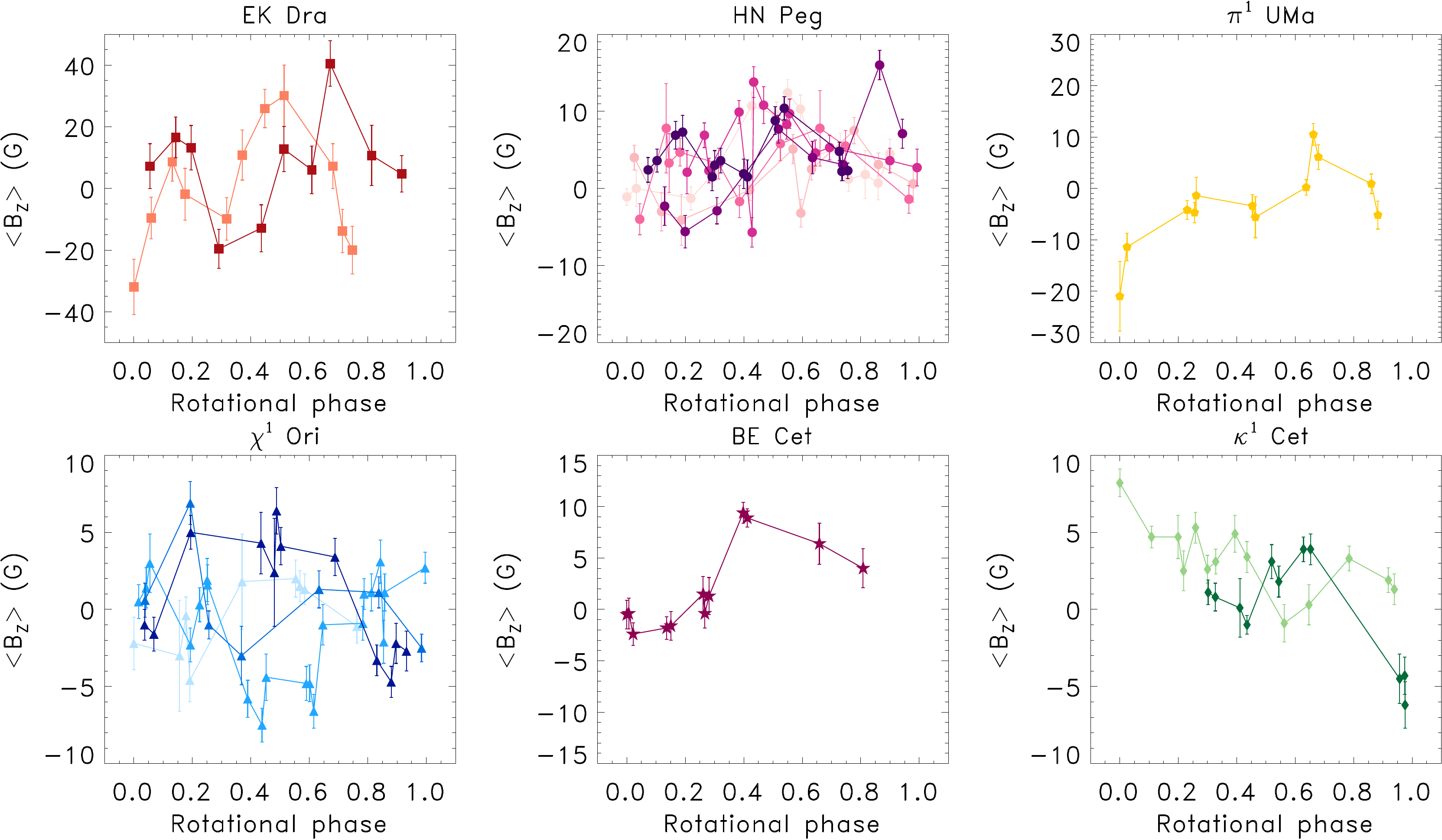} 
\caption{Mean longitudinal magnetic field,  \bz , as a function of rotational phase. Each star is represented by different symbols and each observational set is represented by different colour shades where the earlier observations have a lighter shade. }
\label{bz}
\end{figure*}

\section{Multi-line analysis}
\label{multi}
Even though these stars are expected to be more active than the Sun, it was still not possible to detect any clear polarisation signatures in individual lines in the circular polarisation, Stokes $V$, spectra. The S/N could, however, be increased by applying the multi-line technique called least-squares deconvolution \citep[LSD,][]{Donati97}. LSD is a commonly used technique where the basic assumption is that the entire observed spectrum can be represented by a mean profile convolved with a line mask. This mean profile is effectively a weighted combination of thousands of spectral lines giving it a significantly higher S/N compared to individual spectral lines. Here we have used the LSD code developed by \citet{Kochukhov2010}.

The line specific weights depend on depth for Stokes $I$ and depth, wavelength and effective Land\'{e} factor for Stokes $V$. We used the Vienna Atomic Line Database \citep[{\sc vald},][]{Piskunov1995, Kupka1999} together with {\sc marcs} model atmospheres \citep{Gustafsson2008} to obtain the necessary LSD mask line data for the appropriate stellar parameters. All the {\sc marcs} models that we used had a surface gravity of $\log g$=\,4.5 and a microturbulent velocity of $\xi_{\rm t}=2.0$~km\,s$^{-1}$. Since the {\sc marcs} models have steps of 250~K, we used either a $T_{\rm eff}$ of 5750~K or 6000~K depending on what value was closest to the $T_{\rm eff}$ of the star. All stars were assumed to have a solar metallicity except for BE~Cet for which we used [M/H]=\,$0.22$ \citep{Valenti2005}.  

The LSD profiles were calculated using different wavelength ranges for the two spectropolarimeters. We used lines between 4000--8900~\AA\ and 4000--6865~\AA\ for NARVAL and HAPRSpol respectively. Within these intervals we excluded all lines with an intrinsic depth less than 20\% of the continuum, and lines with a shape deviating from the ``average'' line, for example H$\alpha$ and the Na-doublet. This resulted in line masks consisting of about 2200--3200 lines. Since the two spectrographs have different resolution, we also calculated the LSD profiles on different velocity grids, 1.6~\kms \ and 1.8~\kms \ for HARPS and NARVAL respectively. The associated mean uncertainties per velocity bin can be found in column 5 in Table~\ref{tab_obs}. The application of LSD resulted in secure detections of a magnetic field in the majority of the observations.

From the first moment of Stokes $V$ the mean longitudinal field strength, \bz , can be calculated \citep{Kochukhov2010}. This integral quantity could be zero even though there is a clear magnetic signature in the Stokes V profile itself. This happens when the profile is symmetric, i.e. if the two sides of the star are dominated by opposite and equally strong polarities. Of course, the polarisation signal would also cancel and the Stokes $V$ profile would appear flat if two equally strong magnetic features with opposite polarities were positioned on the same vertical strip perpendicular to the line of sight and close to the equator on opposite hemispheres. Nevertheless, the mean longitudinal magnetic field is still a common and useful diagnostics. All the \bz \ values for the stars in this study can be found in column 6 of Table~\ref{tab_obs} and plotted as a function of rotational phase in Fig.~\ref{bz}. As can be seen in Fig.~\ref{bz} the variation of \bz \ suggests the magnetic field topologies are complex for all the stars.  

\section{Zeeman-Doppler imaging}
\label{zdi}

Each point in a line profile corresponds to an interval of Doppler shifts across the stellar surface, approximately aligned with stellar longitudes. A single Stokes profile can, hence, be used to retrieve longitudinal information of the stellar surface. If there is a time series of Stokes  profiles obtained at different rotational phases, it is also possible to get \textit{latitudinal} information. If the star has a surface inhomogeneity close to the pole, the corresponding line signature would only appear close to the centre of the profile while a feature close to the equator would appear across the entire profile depending on rotational phase. By combining the latitudinal and longitudinal information an image of the entire stellar surface can be reconstructed. 

This technique was first implemented for Stokes $I$ profiles to reconstruct the brightness, temperature and chemical abundance of the stellar surface and is called Doppler imaging \citep{Vogt1987,Piskunov1990,Kochukhov2004}. The same principle can be used for polarised spectra to reconstruct the magnetic field topology. This technique is called Zeeman Doppler imaging \citep[ZDI,][]{Brown91}. It has been shown that magnetic field and temperature reconstruction should, in general, be done simultaneously since magnetic features can be correlated with a cooler temperature compared to the mean $T_{\rm{eff}}$ of the star \citep{Rosen12}. Otherwise the intensity decrease can be interpreted as a lower magnetic field strength. 

Using only circular polarisation to reconstruct the magnetic field is not optimal. Stokes $V$ carries information about the line-of-sight component of the magnetic field vector and is independent of the azimuth angle. The same Stokes $V$ profile can therefore correspond to different field configurations \citep[e.g.][]{Piskunov02}. In the four Stokes parameter study of II~Peg by \citet{Rosen2015} it was shown that Stokes $V$ is not sensitive to complex field structures and that a larger fraction of the magnetic energy is reconstructed in more complex features when linear polarisation is taken into account. It was also shown that the magnetic field can be underestimated in strength, especially the meridional component, when only Stokes $V$ is used. This also agrees with ZDI experiments \citep{Rosen12}. Nevertheless, we expect Stokes $V$ inversions to provide some basic information about the global magnetic field topology of the target stars.

In order to model the observed LSD profiles we have assumed that their behaviour could be described by a single spectral line with average line parameters. The local model Stokes profiles were computed by solving the polarised radiative transfer equation analytically by using the Milne-Eddington approximation. The central wavelength and effective Land\'{e} factor was taken from the LSD line mask, the shape of the line was set by a combination of a depth parameter and by a Voigt function characterised by the two broadening parameters. The linear limb-darkening coefficient was established from theoretical continuum spectral calculations for the mean wavelength. 

It has been shown that LSD Stokes $IV$ profiles can be accurately reproduced by a single-line calculation if the magnetic field is weak, $\leq$2~kG, \citep{Kochukhov2010}. Since these stars are expected to have global fields significantly weaker than 2~kG and since we are not dealing with linear polarisation, which has been shown not to be reproducible by a single-line calculation for any field strength, the single-line approach should be adequate. This approach has been used in several other cool star studies, \citep[e.g.][]{Fares2012,Donati14,Hussain2016}.
 
We reconstructed the magnetic field of all stars, and the brightness distribution for one star, EK~Dra, which is the only object in our sample showing rotational modulation of Stokes $I$ profiles. Even though the distortions in Stokes $I$ were quite small, especially in the observational set form 2012, it was still possible to do brightness inversions for EK~Dra. The brightness distribution was taken into account in the magnetic field reconstruction of this star.     

The magnetic field topology is expressed in terms of spherical harmonics by specifying the harmonic expansion coefficients $\alpha_{l,m}$, $\beta_{l,m}$, $\gamma_{l,m}$, where $l$ is the angular degree and $m$ is the azimuthal order of each mode. The coefficients $\alpha_{l,m}$ represent the radial poloidal field component, $\beta_{l,m}$ the horizontal poloidal component and $\gamma_{l,m}$ the horizontal toroidal component. The maximum angular degree was set to $l_{\rm max}=10$ for all stars since less than 1\% of the total magnetic energy is contained in $l=10$.

Since Stokes $IV$ or Stokes $V$ inversions are ill-posed problems, we have also applied a regularisation. For the magnetic field, we applied a regularisation function to suppress high order modes, $\sum_{l,m}l^2(\alpha_{l,m}^2+\beta_{l,m}^2+\gamma_{l,m}^2)$. For the brightness inversions we used Tikhonov regularisation to smooth out the difference in brightness between neighbouring surface areas. We also used the following penalty function, $R$, with empirically adjusted regularisation parameter $\Lambda$ 

$$
R = \left\{ \begin{array}{ll}
\sum_{i}\Lambda C_1(T_i - T_0)^2, &\mbox{ if $T_i \le T_0$} \\  
\sum_{i}\Lambda C_2(T_i - T_0)^2, &\mbox{ if $T_i > T_0$} 
       \end{array} \right. 
$$
where constants $C_1$ and $C_2$ are set to 1 and 10 respectively, and $T_0=1$. This penalty function sets the mean value of brightness distribution and suppresses the brightness values larger than 1 more strongly than values smaller than 1.

\section{Results}

\begin{table}[t!]
\caption{Phase coverage for each observational set and corresponding goodness of fit with respect to average S/N value of the LSD profiles.}
\label{covfit}
\centering
\begin{tabular}{ccccc}
\hline\hline
Star                & Obs.       & Phase coverage       &  $E_{\rm fit}$    &  $E_{\rm fit}$  \\
name              &   epoch                      & (\%)                           &  Stokes $V$       &  Stokes $I$  \\
\hline
EK~Dra          & 2007.1 & 51 & 1.40 & 2.37  \\
EK~Dra          & 2012.1 & 50 & 1.52 & 1.10  \\
	 
HN~Peg        & 2007.6 & 62 &  1.18 &  - \\
HN~Peg        & 2008.6 & 46 & 1.16 & -  \\
HN~Peg        & 2009.5 & 45& 1.28 &  -\\
HN~Peg        & 2010.5 & 51 &1.60  &  - \\
HN~Peg        & 2011.5 & 40 & 1.04  & - \\
HN~Peg        & 2013.7 & 46 & 1.23 &  - \\

$\pi^1$~UMa  & 2007.1  & 41 & 1.30 & - \\
				  
$\chi^1$~Ori  & 2007.1 & 32 & 1.03 & - \\
$\chi^1$~Ori  & 2008.1 &  59 & 1.74 & - \\
$\chi^1$~Ori  & 2010.8 & 35 & 1.42 &  -\\
$\chi^1$~Ori  & 2011.9 & 45 & 1.34 & - \\
				  
BE~Cet     &2013.7       & 45 & 1.17 & -\\
				  
$\kappa^1$~Cet & 2012.8    & 59 & 1.44 &- \\
$\kappa^1$~Cet & 2013.7    & 37 & 1.10 &- \\
			
\hline
\end{tabular}
\end{table}

\subsection{Line profiles and surface maps}
\begin{figure*}
\centering
\includegraphics[width=\textwidth]{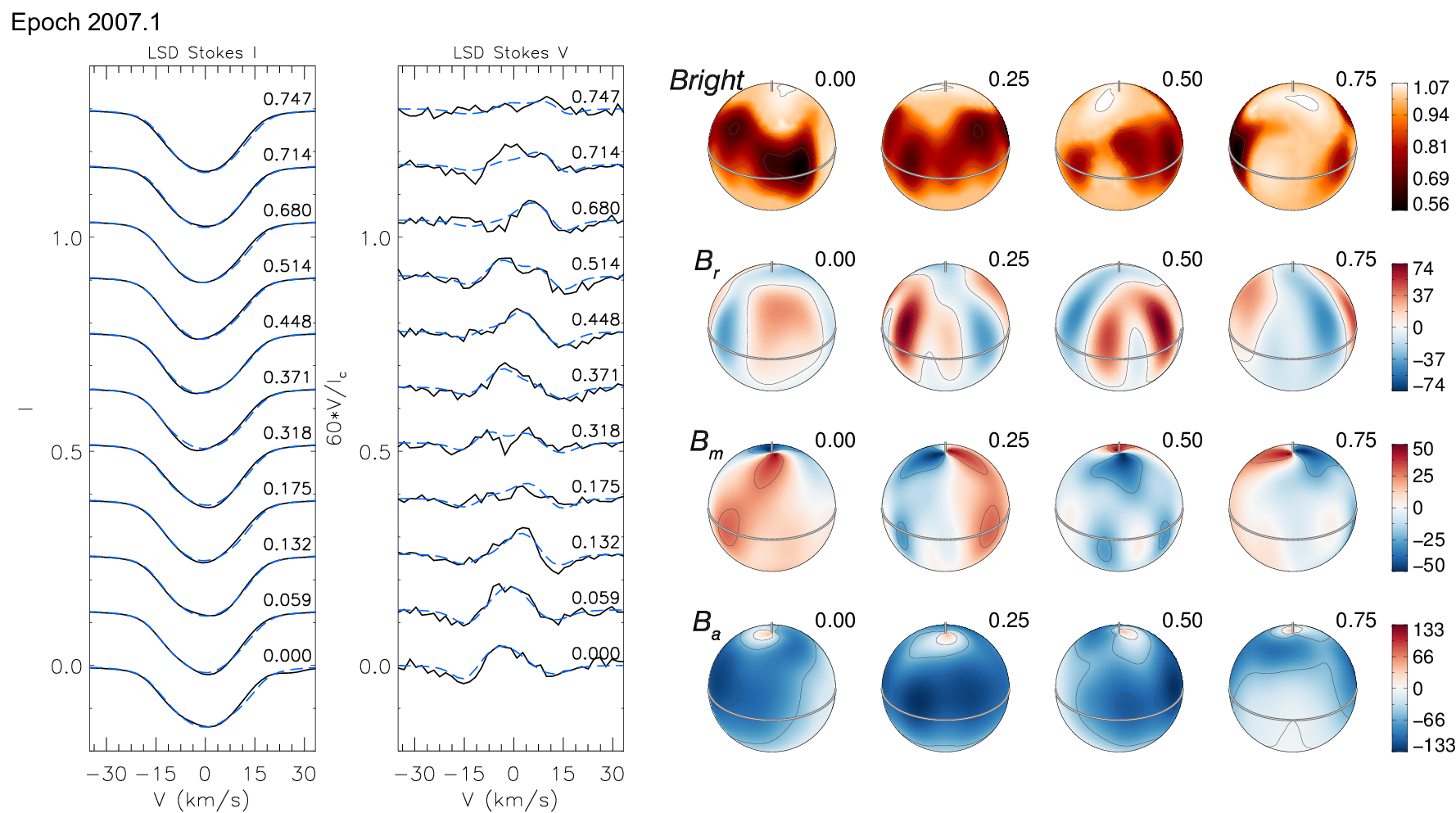} 
\includegraphics[width=\textwidth]{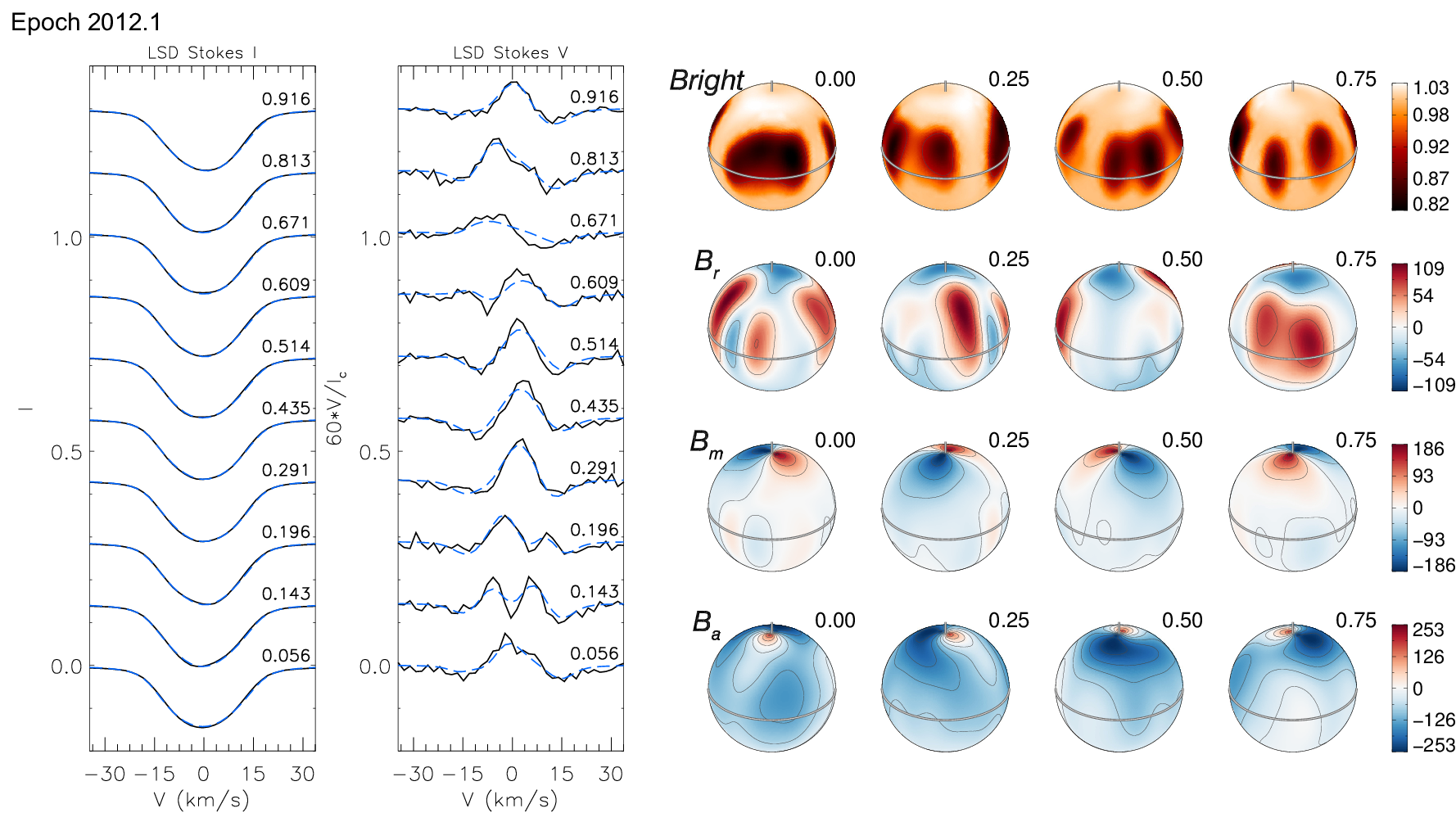}
\caption{Line profiles and reconstructed brightness and magnetic field distribution of EK~Dra. The observed LSD profiles are represented by the black solid lines and the corresponding model profiles are represented by the blue dashed lines. Both the Stokes $I$ and $V$ profiles have been shifted vertically and the Stokes $V$ profiles have been magnified by a factor of 60. Rotational phases are indicated next to each profile. The stellar surface is shown at two different rotational phases. The brightness distribution is shown in the top row of spherical maps; the radial, meridional and azimuthal field distributions are shown in the following three rows. The magnetic field strength is given in G. The upper set of profiles and maps corresponds to the 2007.1 epoch and the lower to the 2012.1 epoch. }
\label{ek-map}
\end{figure*}

\begin{figure*}
\centering
\includegraphics[width=\textwidth]{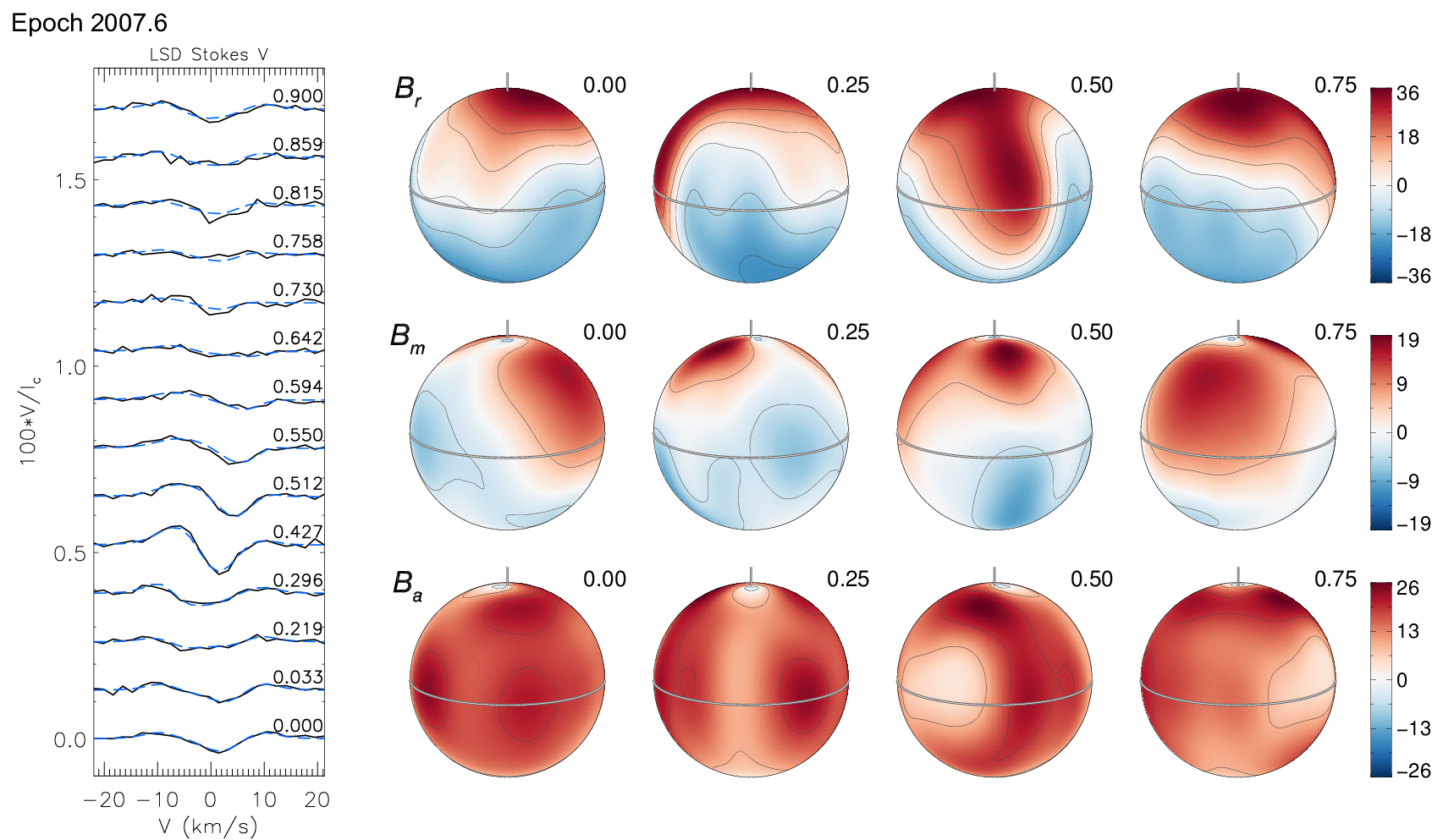}  
\includegraphics[width=\textwidth]{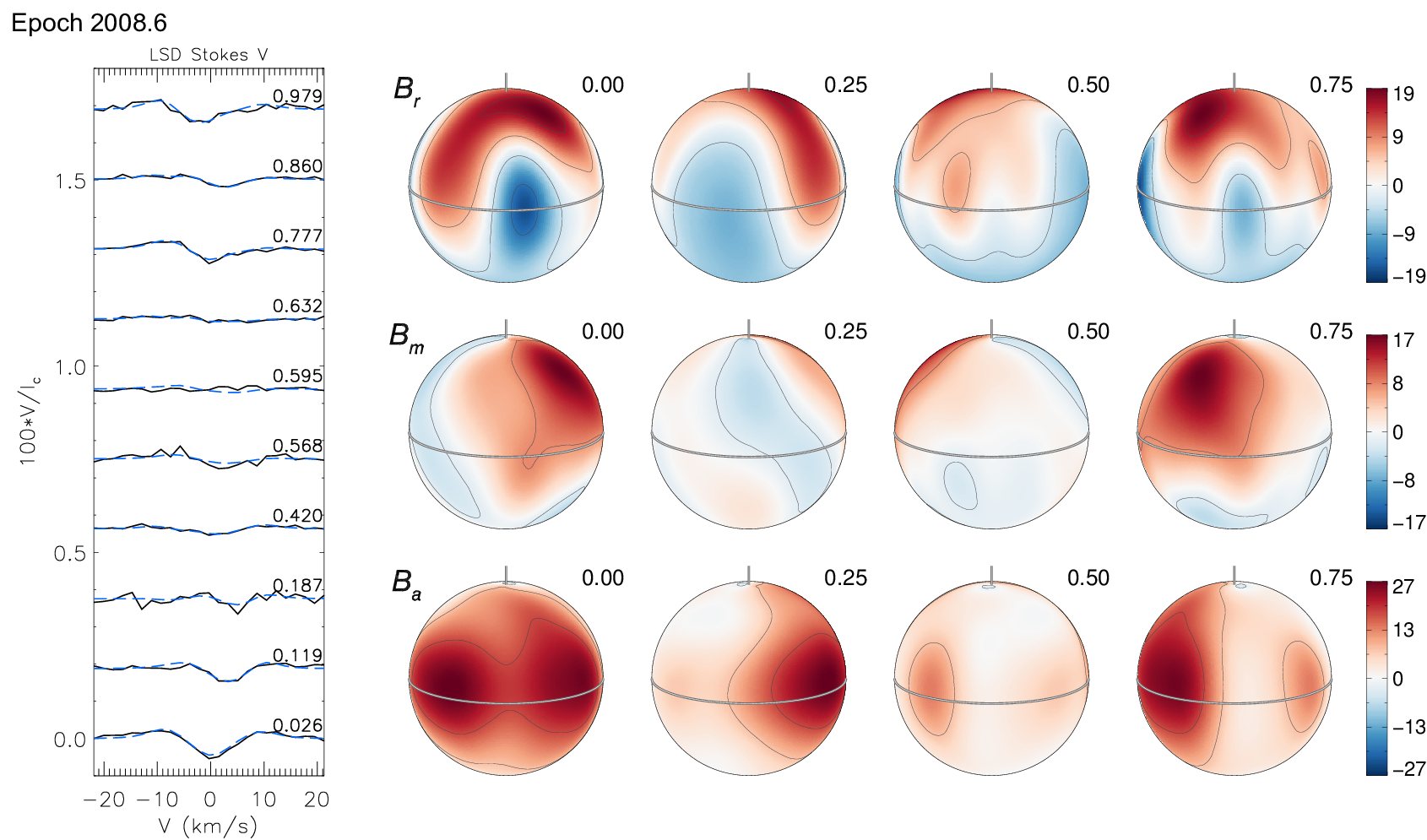} 
\caption{Same as Fig.~\ref{ek-map} but here the line profiles and maps correspond to ZDI reconstructions for HN~Peg. Only Stokes $V$ was used to reconstruct the magnetic field and the Stokes $V$ profiles have been magnified by a factor of 100. The upper set of profiles and maps corresponds to the 2007.6 epoch, and the lower to the 2008.6 epoch.}
\label{hn-map1}
\end{figure*}

\begin{figure*}
\centering
\includegraphics[width=\textwidth]{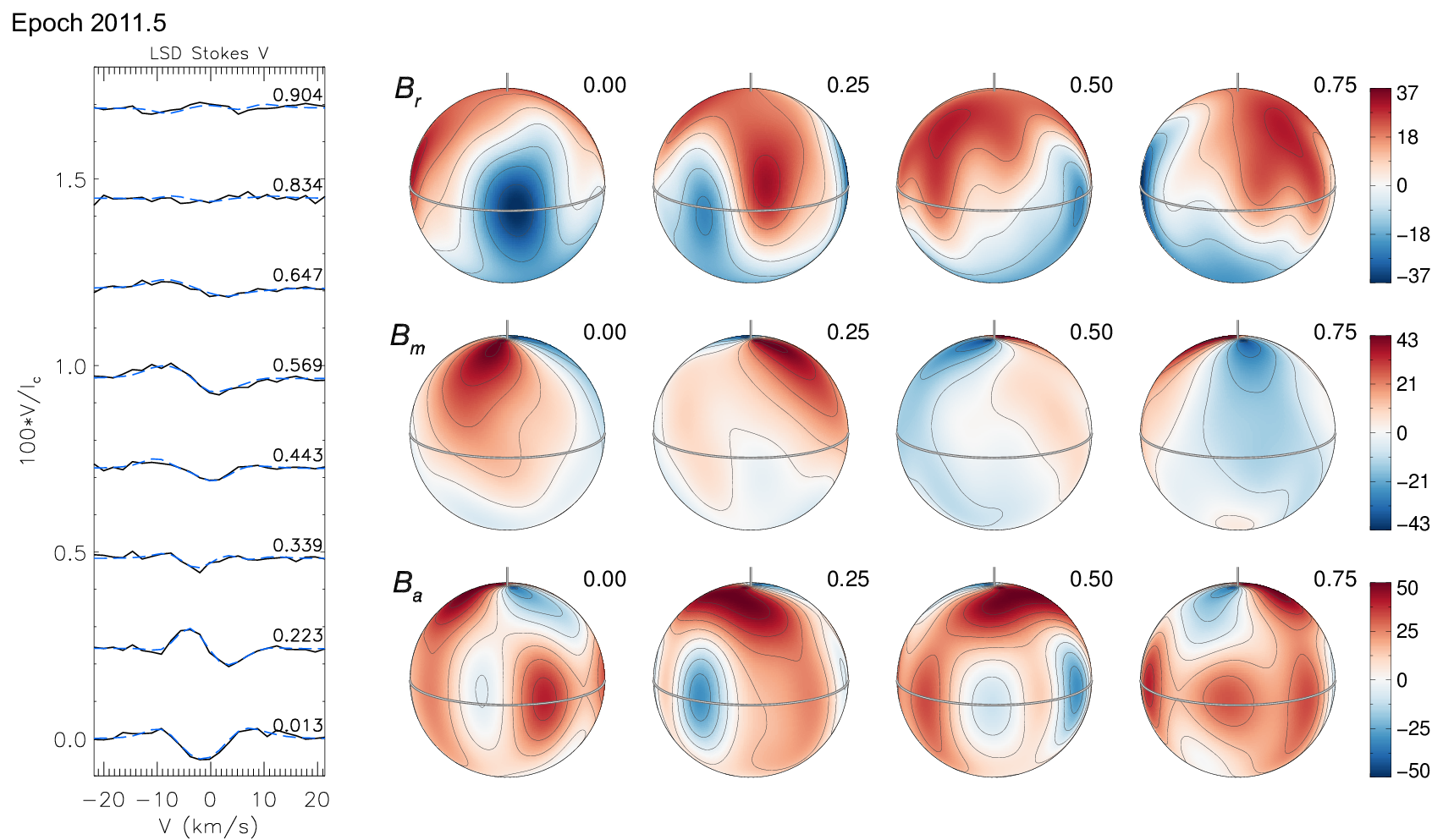}  
\includegraphics[width=\textwidth]{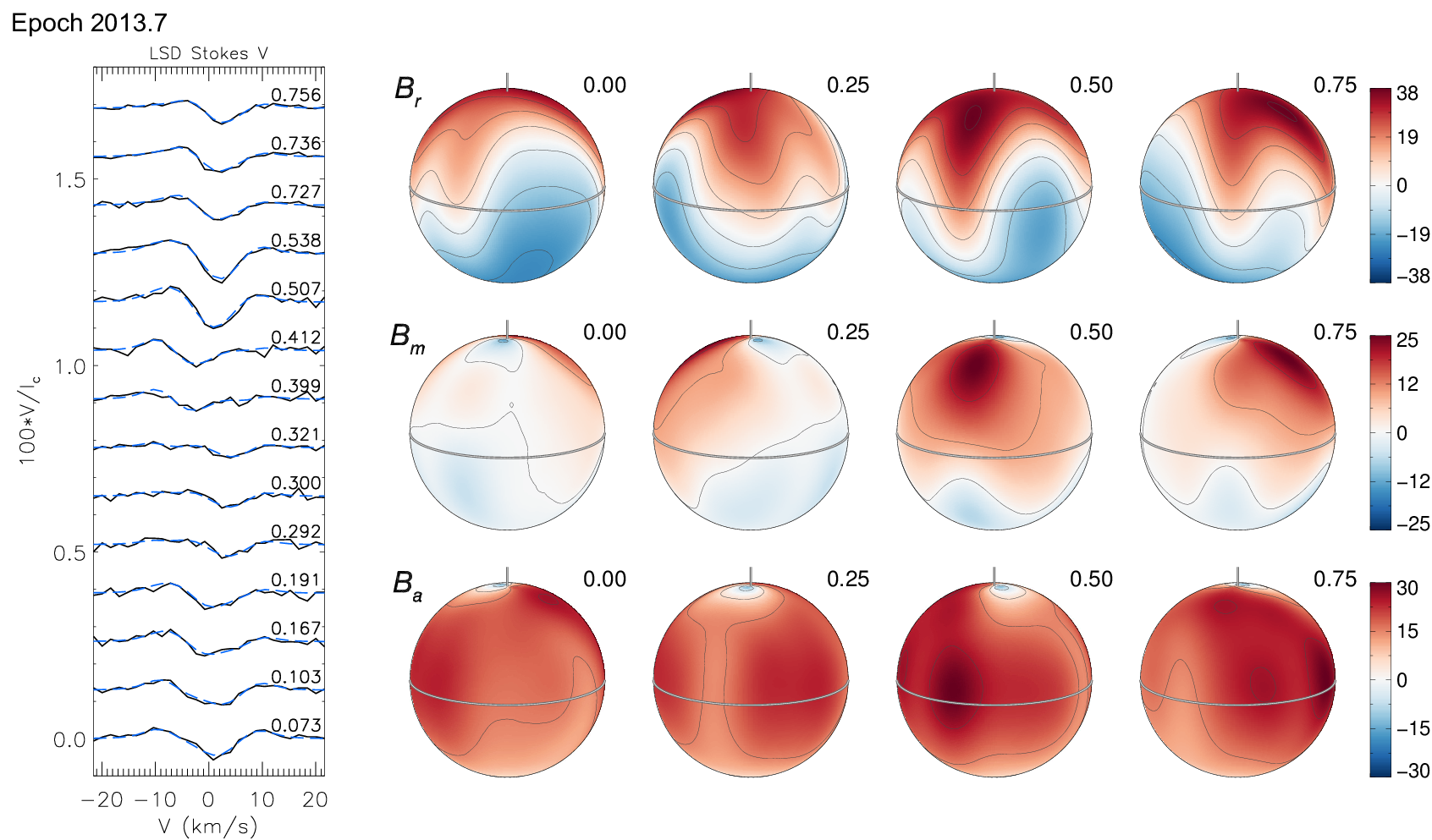} \\
\caption{Same as Fig.~\ref{hn-map1} but here the line profiles and maps correspond to ZDI reconstructions for HN~Peg from the epoch 2011.5 epoch and the 2013.7 epoch.}
\label{hn-map}
\end{figure*}

\begin{figure*}
\centering
\includegraphics[width=\textwidth]{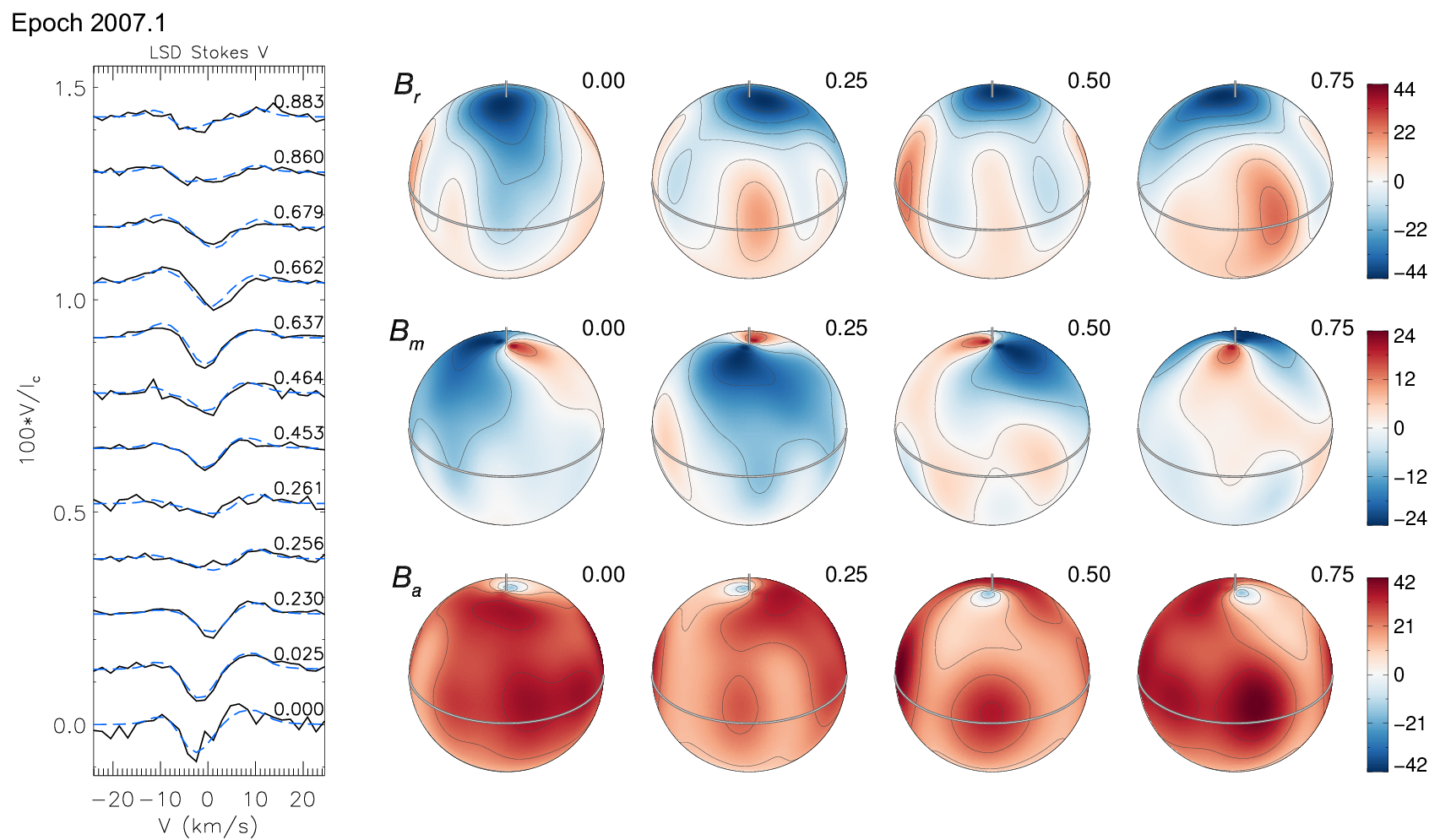}  
\caption{Same as Fig.~\ref{hn-map1} but here the line profiles and maps correspond to the observations of $\pi^1$~UMa from the 2007.1 epoch. }
\label{pi-map}
\end{figure*}

\begin{figure*}
\centering
\includegraphics[width=\textwidth]{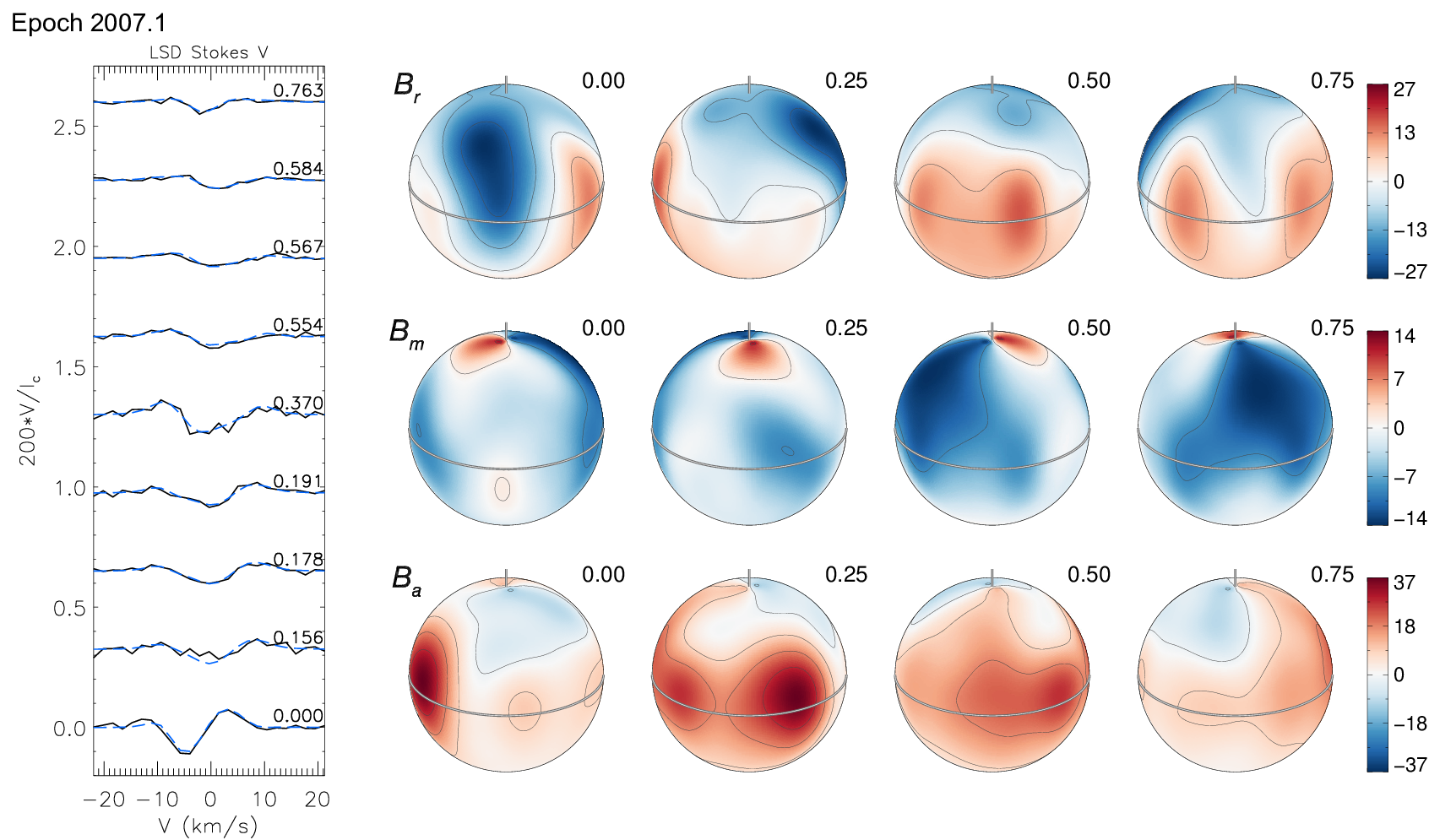}  
\includegraphics[width=\textwidth]{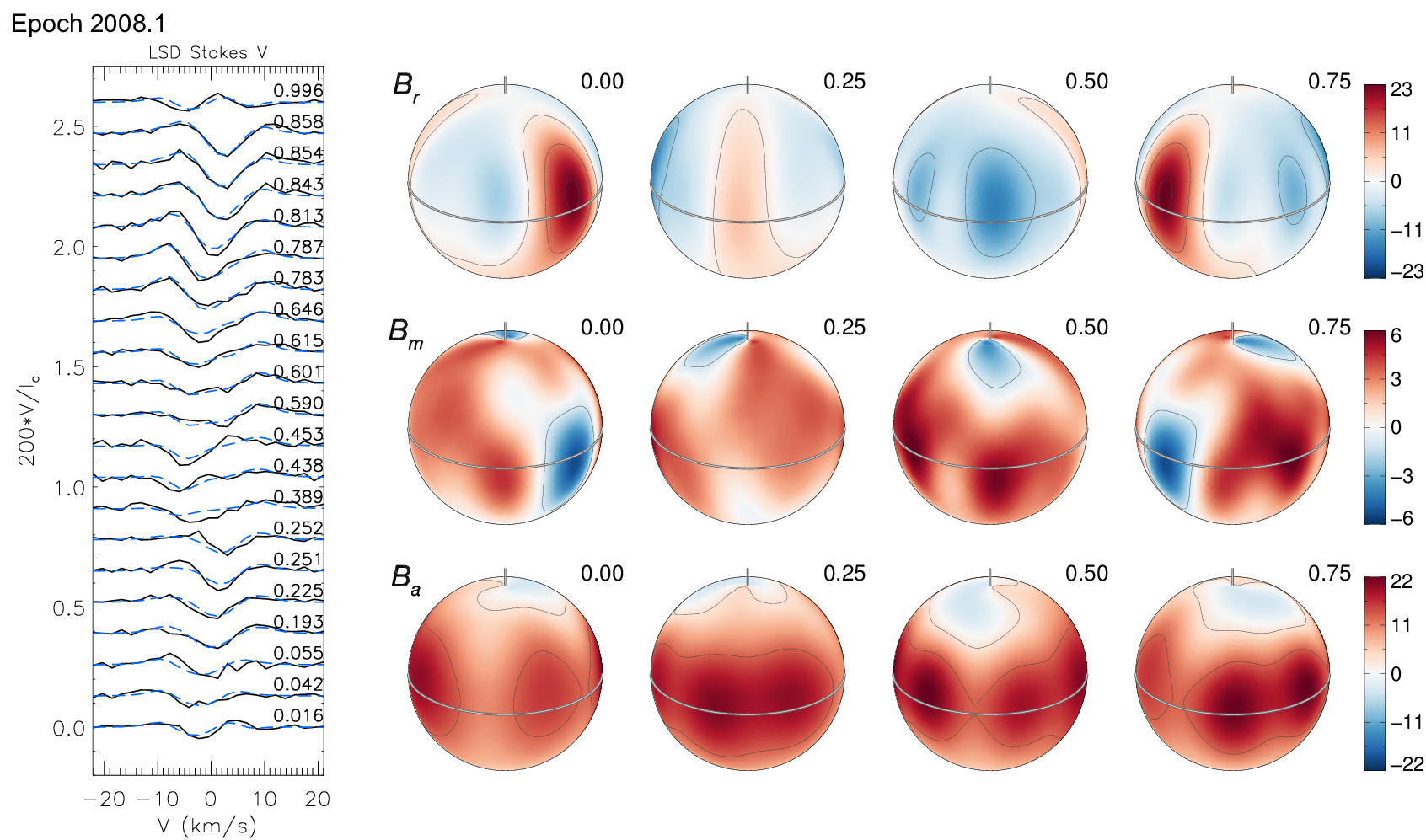}  
\caption{Same as Fig.~\ref{hn-map1} but here the line profiles and maps correspond to the observations of $\chi^1$~Ori. The Stokes $V$ profiles have been magnified by a factor of 200. The upper set of profiles and maps corresponds to the 2007.1 epoch, the upper right to the 2008.1. }
\label{chi-map1}
\end{figure*}

\begin{figure*}
\centering
\includegraphics[width=\textwidth]{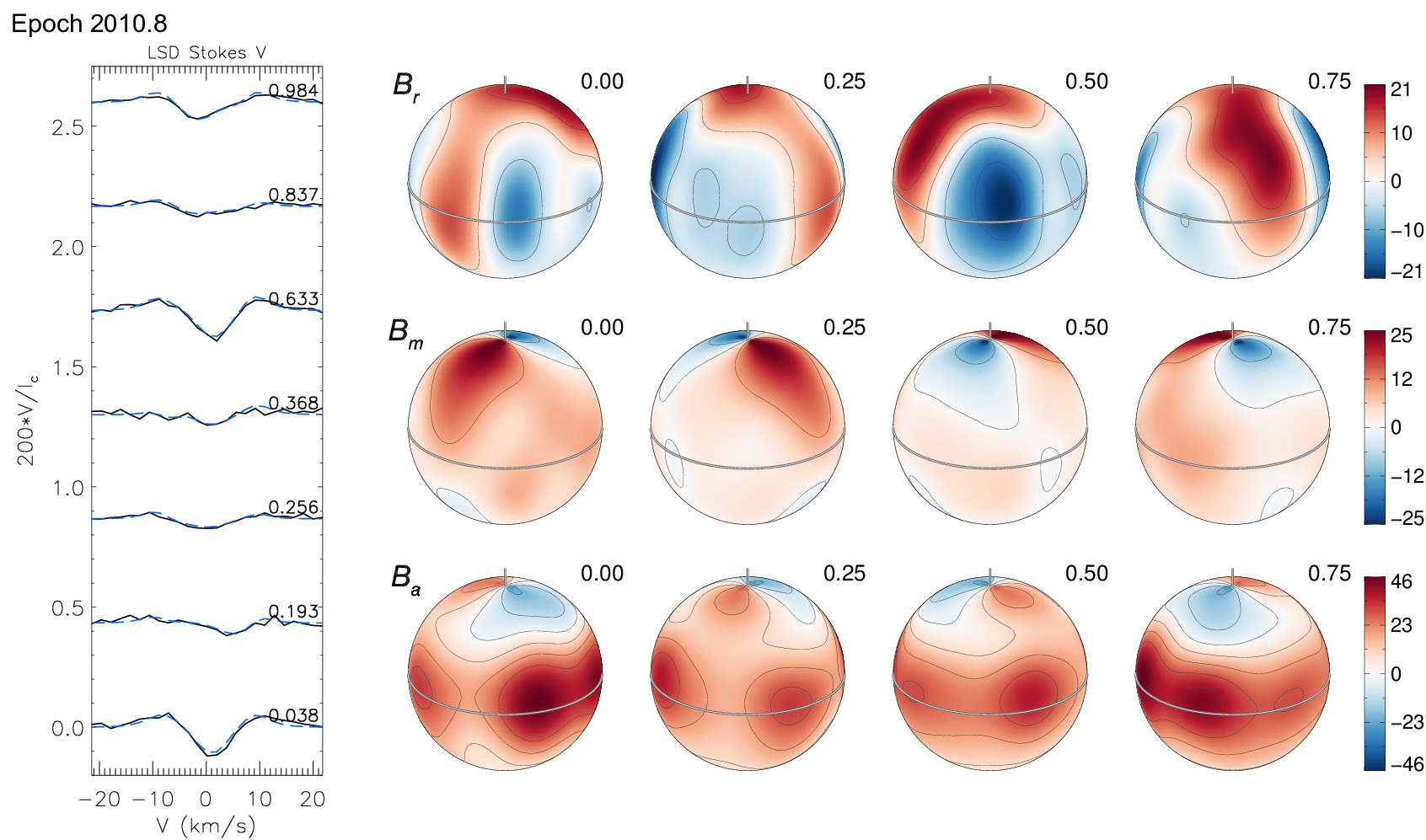}
\includegraphics[width=\textwidth]{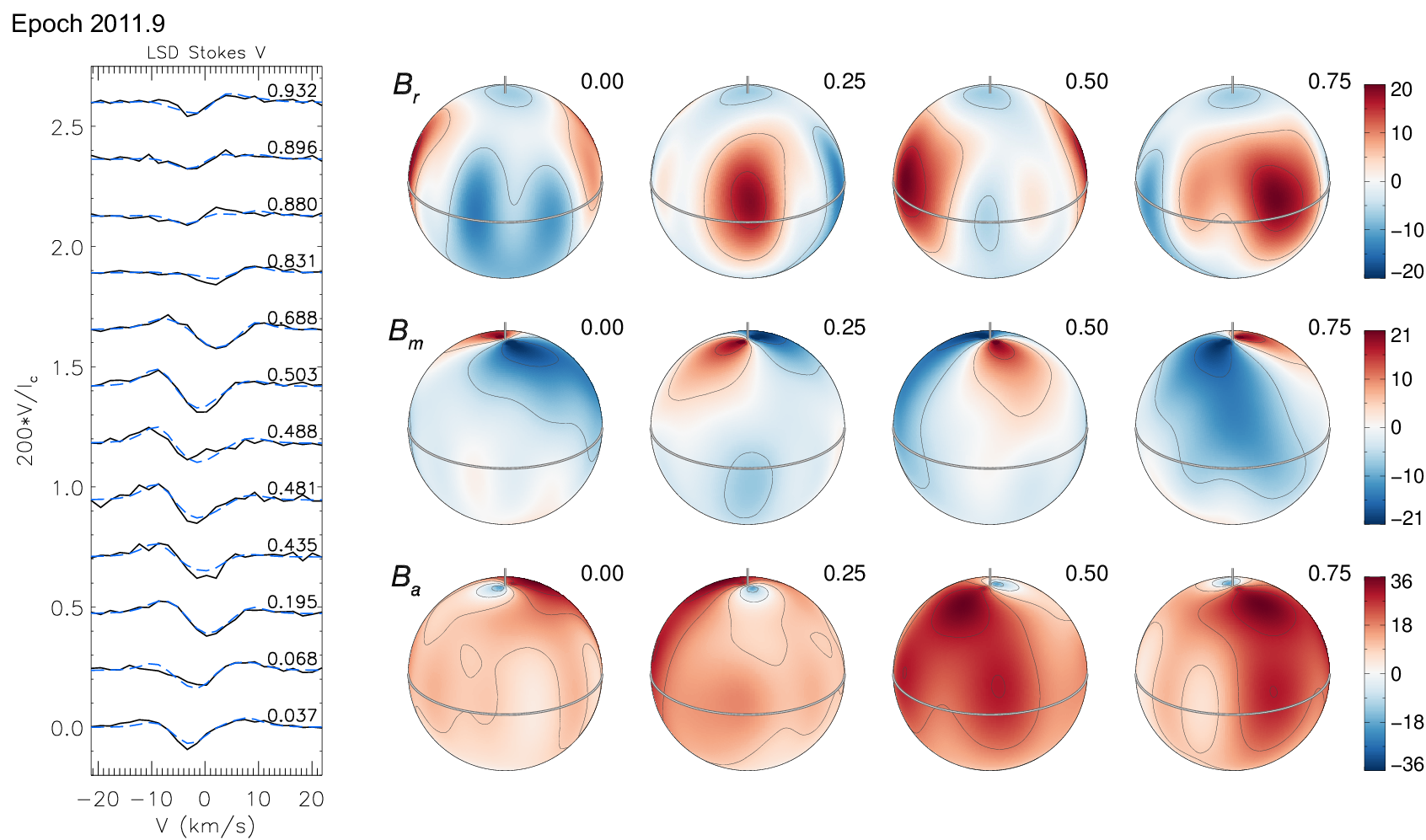}
\caption{Same as Fig.~\ref{chi-map1} but here the line profiles and maps correspond to ZDI reconstructions for $\chi^1$~Ori from the epoch 2010.8 epoch and the 2011.9 epoch }
\label{chi-map}
\end{figure*}

\begin{figure*}
\centering
\includegraphics[width=\textwidth]{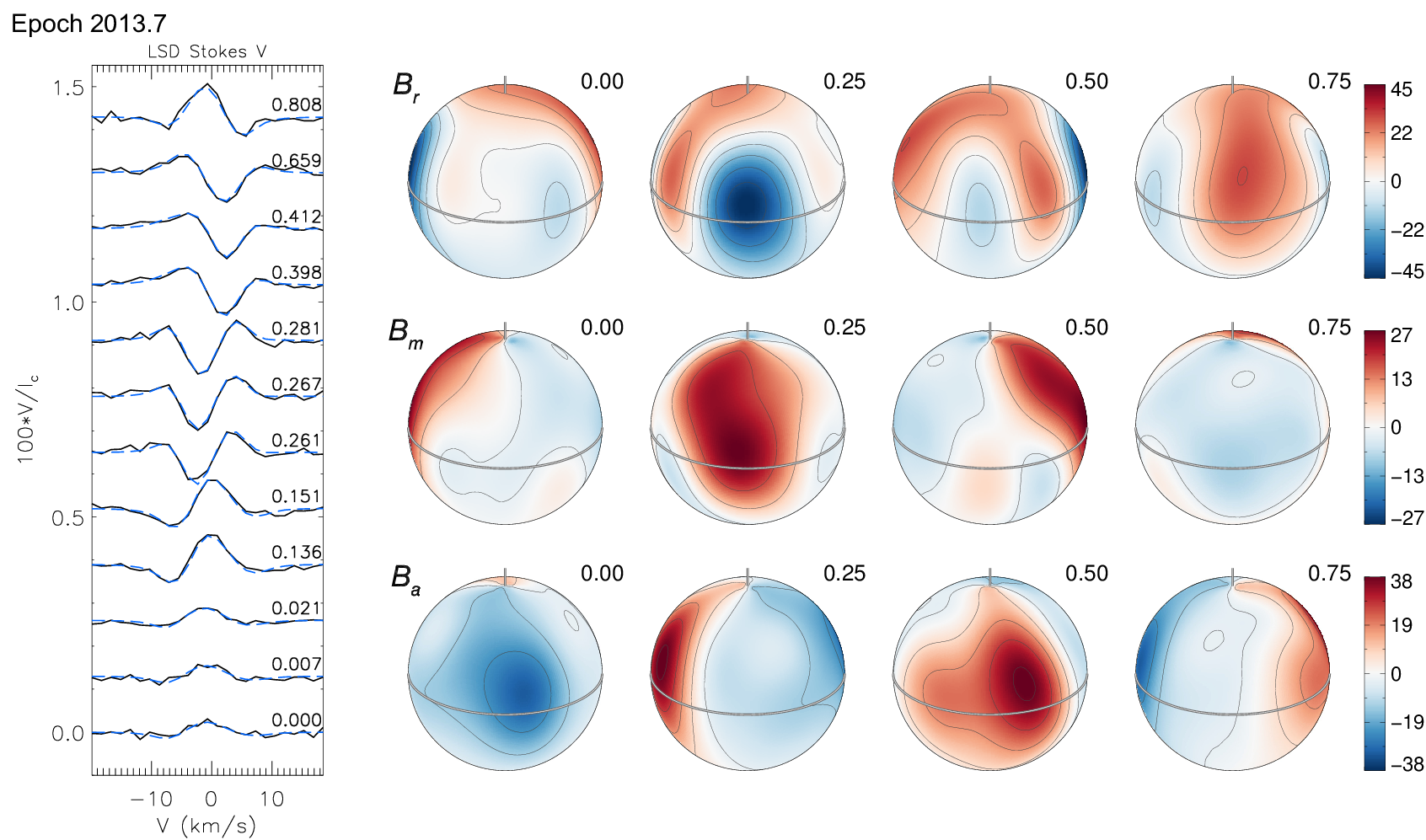}  
\caption{Same as Fig.~\ref{hn-map} but here the line profiles and maps correspond to the observations of BE~Cet from the 2013.7 epoch.}
\label{be-map}
\end{figure*}

\begin{figure*}
\centering
\includegraphics[width=\textwidth]{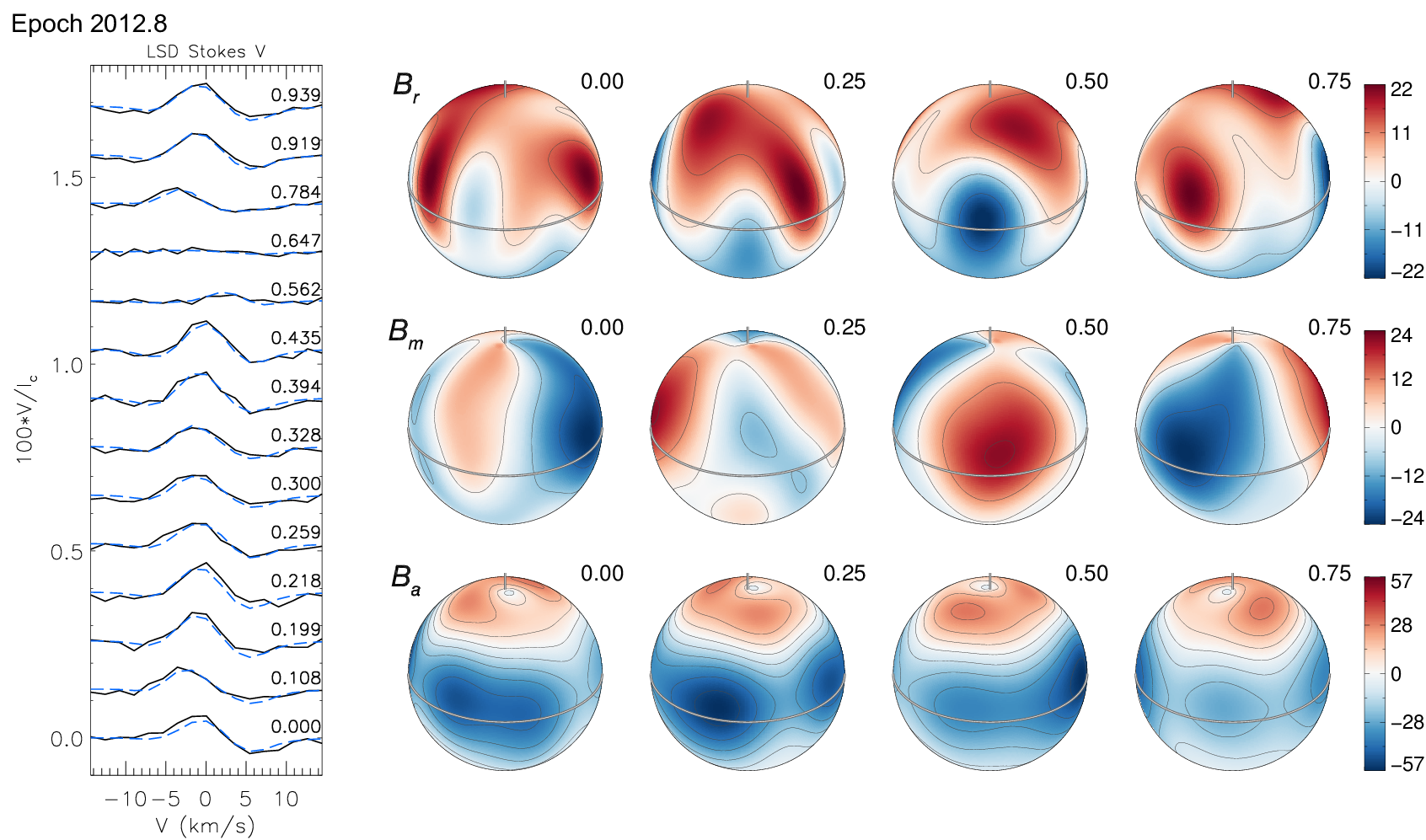} 
\includegraphics[width=\textwidth]{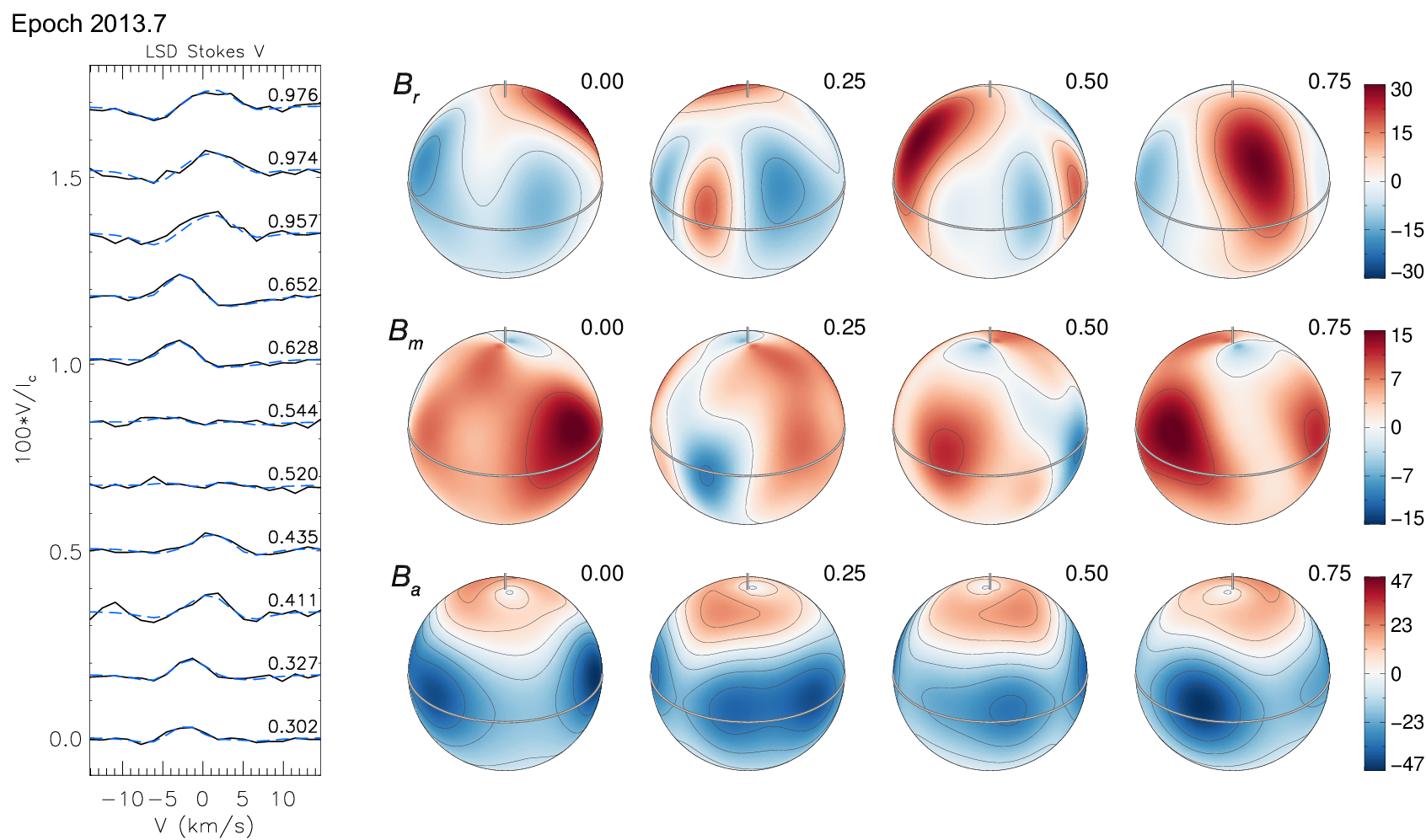}  
\caption{Same as Fig.~\ref{hn-map} but here the line profiles and maps corresponds to the observations of $\kappa^1$~Cet. The left set of profiles and maps correspond to the 2012.8 epoch and the right to the 2013.7 epoch. }
\label{kap-map}
\end{figure*}

The derived surface maps of all stars can be found in Figs.~\ref{ek-map}-\ref{kap-map}. These figures also show a comparison between the observed LSD profiles and the model profiles. In order to get an estimate of the quality and reliability of the maps, we have calculated the phase coverage of each data set assuming that each observation covers 5\% of the rotation period. We have also calculated the ratio between the mean deviation of the observed profiles to the model profiles, and the average error of the observed LSD profiles to get an estimate of the goodness of the fit, $E_{\rm fit}$. This number can be used to make sure that we are not fitting noise, i.e. if $E_{\rm fit}<1$, while still maintaining a good fit. All values of phase coverage and $E_{\rm fit}$ can be found in Table~\ref{covfit}.

Being the youngest star with the shortest rotation period, EK~Dra also showed the strongest magnetic field with a local maximum field strength (by absolute value) of 133~G and 253~G for the 2007.1 and 2012.1 epochs respectively. The second largest maximum field strength value is 57~G for the 2012.8 epoch of $\kappa^1$~Cet and the lowest value is 23~G for the 2008.1 epoch of $\chi^1$~Ori. There are also large differences for multiple observation epochs of individual stars, for instance, the range of HN~Peg is 27~G to 50~G. It is also possible to compare the local maximum field strengths of each component for the same ZDI map. It turns out that in 9 out of 16 ZDI maps the highest local field strength is found for the azimuthal component while the meridional field has the lowest local maximum strength in 10 out of 16 maps. 

\begin{figure*}
\centering
\includegraphics[scale=0.6]{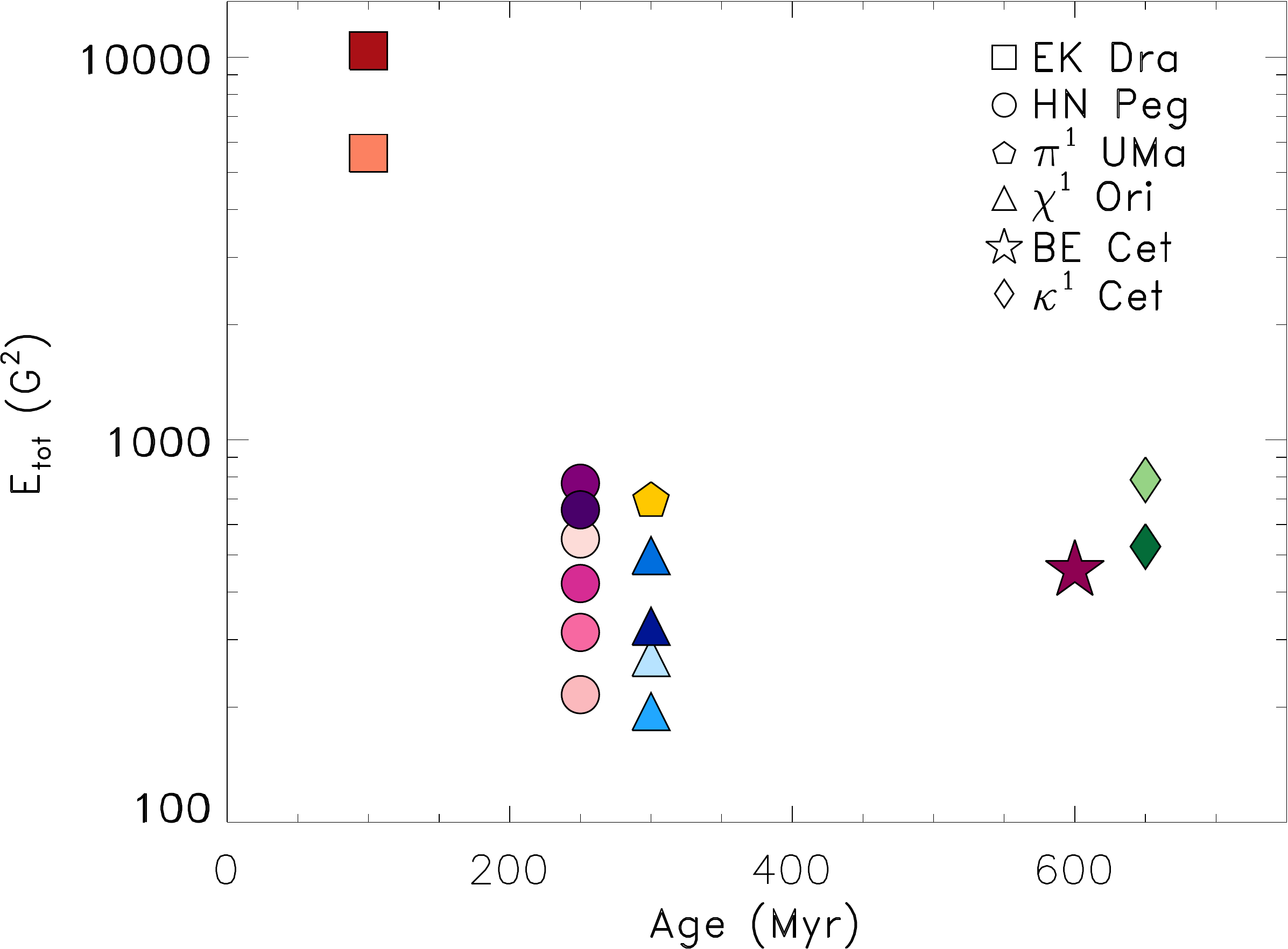} 
\caption{Total magnetic field energy as a function of age. The stars are represented in the same way as in Fig.~\ref{bz}. }
\label{etot}
\end{figure*}

\begin{figure*}
\centering
\includegraphics[scale=0.60]{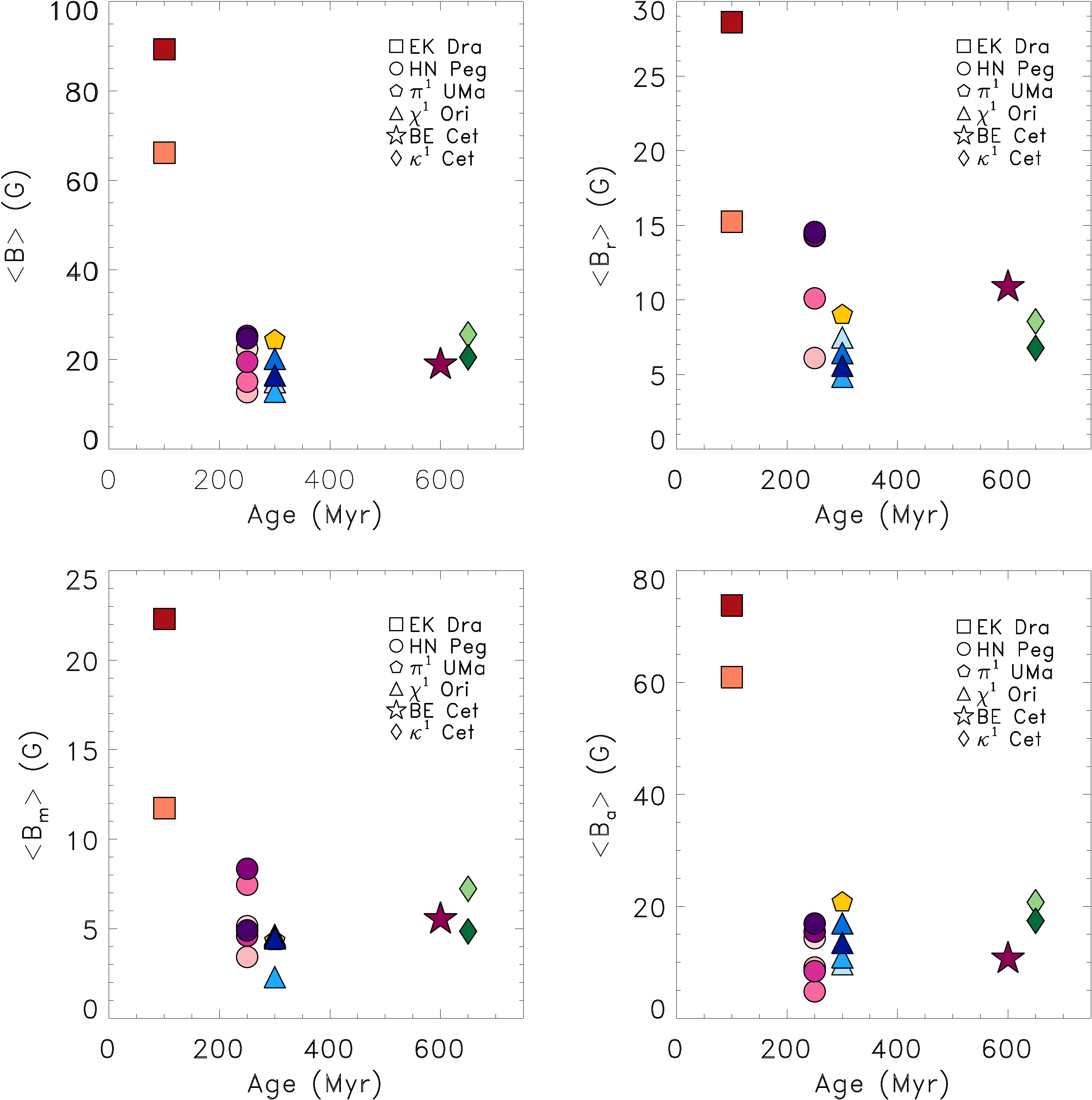} 
\caption{Total mean field strength as a function of age and mean radial, meridional and azimuthal field strengths as a function of age. The stars are represented in the same way as in Fig.~\ref{bz}.}
\label{bvbc}
\end{figure*}

Instead of only looking at the local maximum field strength, it is useful to calculate the total magnetic field energy, $E_{\rm tot}$, which is proportional to $B^2$ integrated over the entire stellar surface. This parameter is plotted in Fig.~\ref{etot} where the different stars are represented by different shapes and different observation epochs by different colours. Another useful parameter is the mean field strength, $\langle B \rangle= \sum_i S^i \cdot \sqrt{(B_{\rm r}^i)^2+(B_{\rm m}^i)^2+(B_{\rm a}^i)^2}$, where $B_{\rm r}^i$, $B_{\rm m}^i$, and $B_{\rm a}^i$ are the local field strengths of the radial, meridional and azimuthal components respectively and $S^i$ is the area of each surface element with a total area normalised to 1 over the 1176 surface elements in total. We have also calculated the mean field strength of each component, $\langle B_{\rm comp} \rangle= \sum_i S^i \cdot |B_{\rm comp}^i|$, where $B_{\rm comp}$ is either $B_{\rm r}$, $B_{\rm m}$ or $B_{\rm a}$. This is useful since the radial field is the most important for the stellar wind \citep{Jardine2013} and because the strengths of the individual components can be compared. Fig.~\ref{bvbc} depicts how $\langle B \rangle$ and $\langle B_{\rm r} \rangle$, $\langle B_{\rm m} \rangle$ and $\langle B_{\rm a} \rangle$ varies with age and all magnetic field strengths are also listed in Table~\ref{mb}. It turns out that $\langle B_{\rm a} \rangle$ is strongest in 13 out of 16 maps while $\langle B_{\rm m} \rangle$ is weakest in 15 out of 16 maps.

\begin{table}[t!]
\caption{Mean magnetic field strength.}
\label{mb}
\centering
\begin{tabular}{cccccc}
\hline\hline
Star                & Obs.       & $\langle B \rangle$       &  $\langle B_{\rm r} \rangle$    &  $\langle B_{\rm m} \rangle$ & $\langle B_{\rm a} \rangle$ \\
name              &   epoch                      & (G)      & (G)       &  (G) & (G)  \\
\hline
EK~Dra         & 2007.1      & 66 & 15 & 12 & 61 \\
EK~Dra          & 2012.1 	  & 89 & 29 & 22 & 74 \\
	 
HN~Peg        & 2007.6       & 22 &  14 &  5 & 14 \\
HN~Peg        & 2008.6 	  & 13 & 6 & 3 & 9 \\
HN~Peg        & 2009.5	  & 15 & 10 &  7 & 5 \\
HN~Peg        & 2010.5 	  & 20 & 14  & 5 & 8 \\
HN~Peg        & 2011.5 	  & 25 & 14  & 8 &  15 \\
HN~Peg        & 2013.7 	  & 25 & 15 &  5 & 17 \\

$\pi^1$~UMa & 2007.1     & 24 & 9 & 4 & 21 \\
				  
$\chi^1$~Ori  & 2007.1 	& 15 & 7 & 5 & 10 \\
$\chi^1$~Ori  & 2008.1 	&  13 & 5 & 2 & 11 \\
$\chi^1$~Ori  & 2010.8 	& 20 & 6 &  5 & 17 \\
$\chi^1$~Ori  & 2011.9 	& 16 & 6 & 4 & 13  \\
				  
BE~Cet     & 2013.7           & 19 & 11 & 6 & 11 \\
				  
$\kappa^1$~Cet & 2012.8    & 26 & 9 & 7 & 21 \\
$\kappa^1$~Cet & 2013.7    & 21 & 7 & 5 & 17 \\
			
\hline
\end{tabular}
\end{table}

\begin{figure*}
\centering
\includegraphics[scale=0.63]{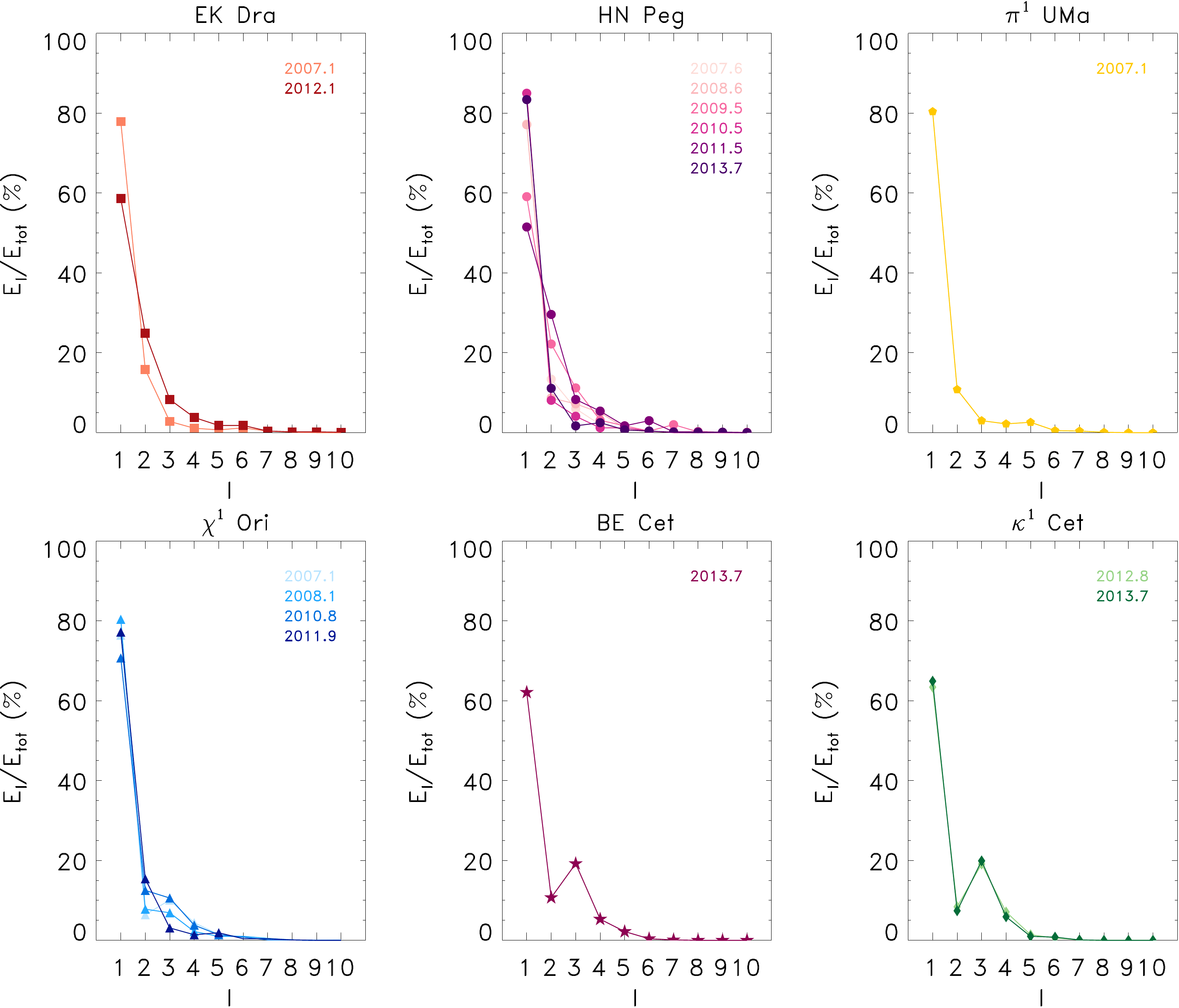} 
\caption{Magnetic energy as a function of $l$ for each star. The stars are represented in the same way as in Fig.~\ref{bz}. }
\label{lmode}
\end{figure*}

\begin{figure*}
\centering
\includegraphics[scale=0.63]{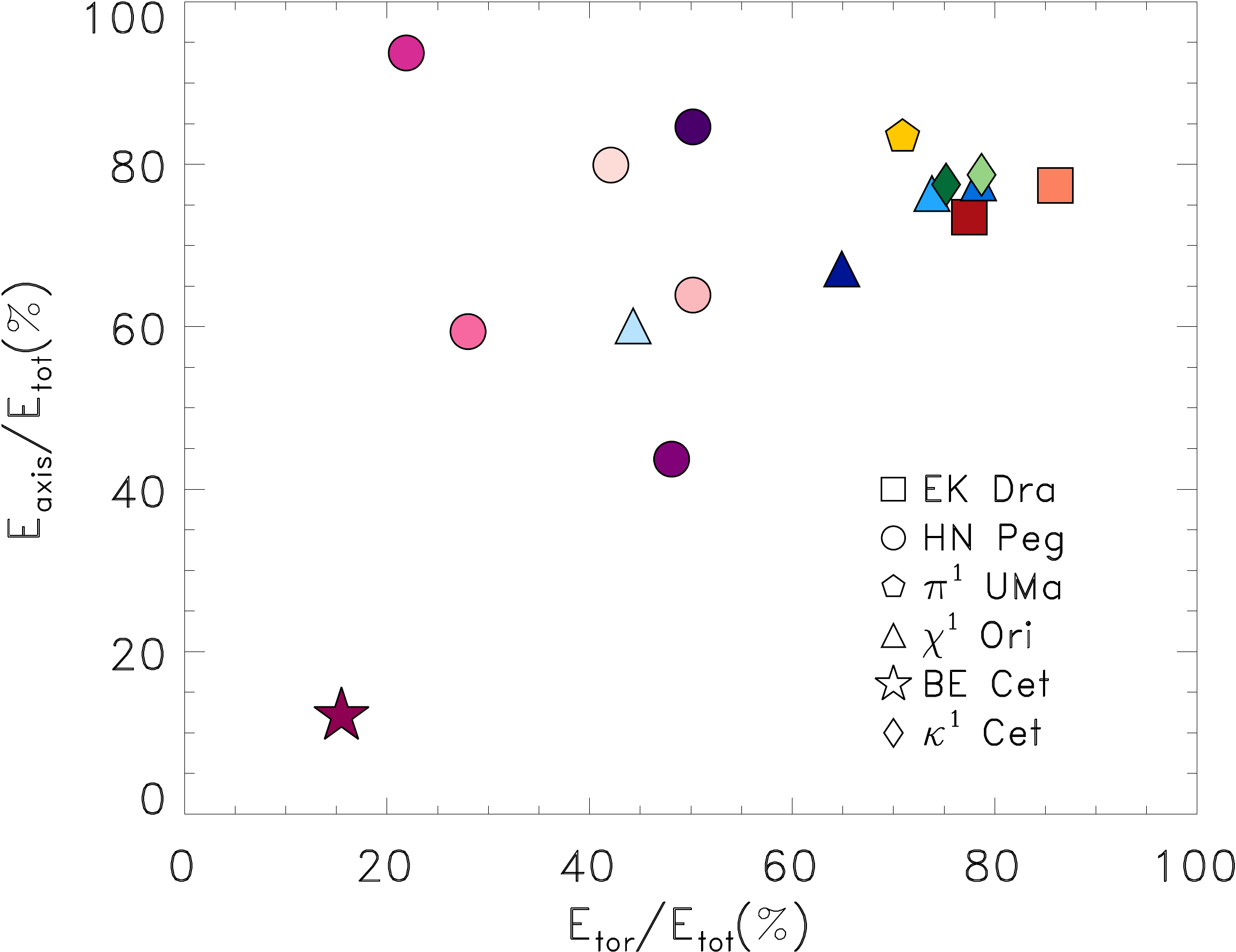} 
\caption{Fraction of axisymmetric field energy to the total field energy as a function of the fraction of toroidal field energy to the total field energy. The stars are represented in the same way as in Fig.~\ref{bz}.}
\label{topo}
\end{figure*}

\begin{figure*}
\centering
\includegraphics[scale=0.63]{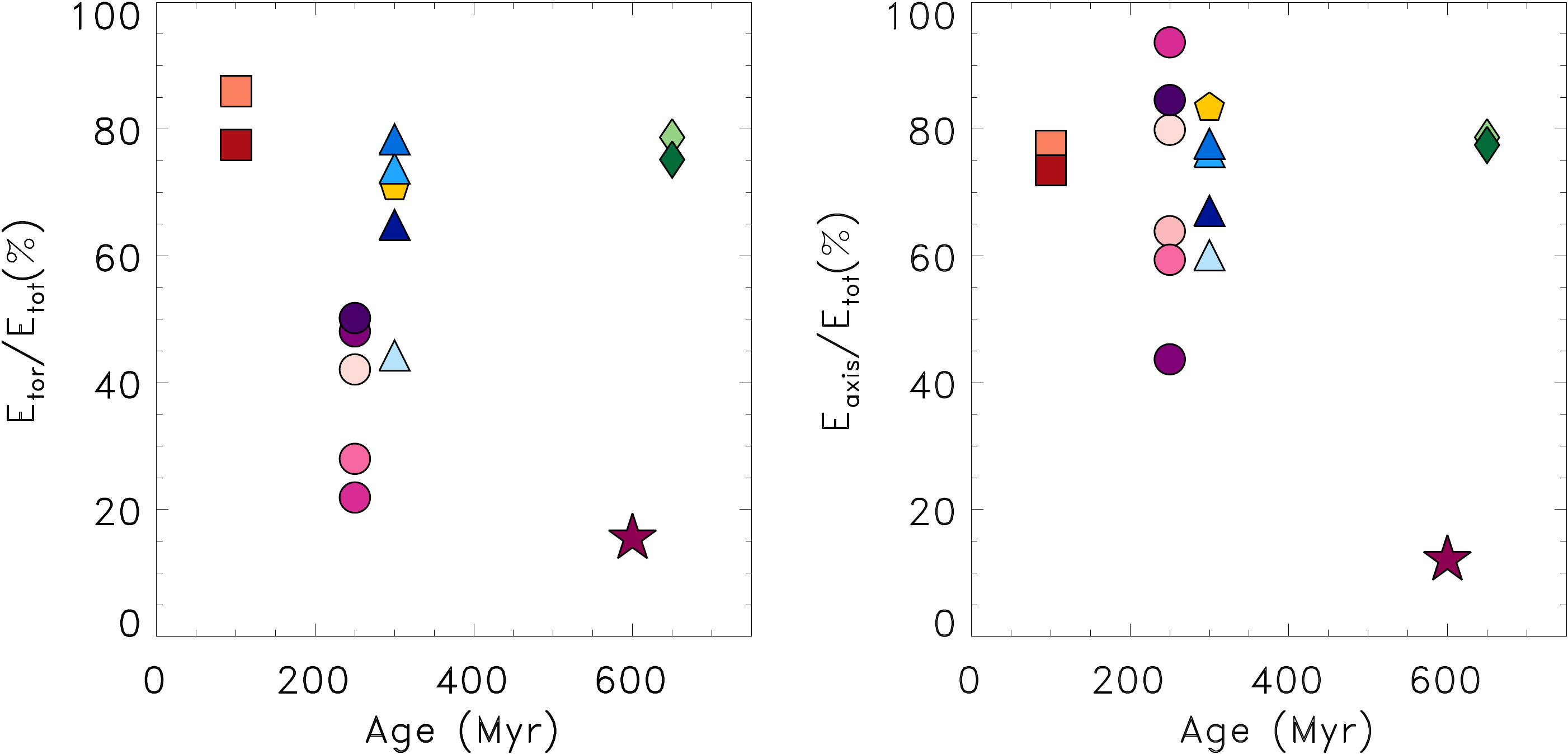} 
\caption{Fraction of toroidal field energy to the total field energy as a function of age and the fraction of axisymmetric field energy to the total field energy as a function of age. The stars are represented in the same way as in Fig.~\ref{bz}.}
\label{avtorsym}
\end{figure*}

The two observation epochs of EK~Dra show a magnetic energy which is 7 and 13 times larger than the highest value for any of the other stars as can be seen in Fig.~\ref{etot}. The highest value for the other five stars is found for the oldest star, $\kappa^1$~Cet although both HN~Peg and $\pi^1$~UMa have similarly high values. The lowest value is found for $\chi^1$~Ori, being about 4 times smaller. However, both HN~Peg and $\chi^1$~Ori exhibit a relatively large span in magnetic energy corresponding to different epochs, with the ratio of highest to lowest of 3.6 and 2.6 respectively. 

For EK~Dra, $\langle B \rangle$ is 66~G and 89~G for the two epochs respectively, which is at least 2.5 and 3.4 times larger than for any other star which have ranging values between 13-26~G. HN~Peg alone has a span of 13-25~G. A similar pattern is seen for $\langle B_{\rm a}\rangle$ but with an even larger discrepancy between EK~Dra and the other stars. The mean radial field strength, $\langle B_{\rm r} \rangle$, is 29~G for the 2012.1 epoch of EK~Dra. This is about twice as large as for any other epoch, including the 2007.1 of EK~Dra which has a similar strength as four of the observation epochs of HN~Peg. Again, the lowest strength, 5~G, is found for $\chi^1$~Ori. A similar pattern is seen for $\langle B_{\rm m} \rangle$ but with slightly lower strengths overall.

$\chi^1$~Ori displays two polarity switches during the four years it was observed as can be seen in Figs.~\ref{chi-map1}-\ref{chi-map}. In the end of January to beginning of February 2007, the visible pole has a negative radial field and the meridional field is dominantly negative with a small positive area close to the pole. A year later, in end the of January to beginning of February 2008, the visible pole has a very weak, only a few G, but positive radial field and the meridional field seems to have the opposite topology compared to the 2007.1 epoch. The pole also has a positive, but stronger, radial field for the period of late September to beginning of October 2010 with a similar structure of the meridional field. Just over a year later, in late October to beginning of November 2011, the polar radial field has switched back to being negative and the meridional map is again similar to the map from the 2007.1 epoch. Since this star was not observed during 2009, two polarity reversals may have been missed and the star could have a magnetic cycle of approximately 2 years. If there was no polarity switch between the 2008.1 and 2010.8 epoch, the magnetic cycle could instead be approximately 6 or 8 years. All the other stars which were observed during multiple epochs showed the same polarity in consecutive ZDI images.

EK~Dra was the only star for which it was possible to do brightness inversions. Both epochs show a band of dark spots along the equator, but with different intensities. We do not see the dark polar cap that was found by \citet{Strassmeier1998}. The strong azimuthal feature along the equator in the 2007 maps seems to overlap with the band of dark spots. This is, however, not the case for the 2012 observations where the strongest magnetic feature is found near the pole.

\subsection{Global magnetic field topology}

Since we are using a spherical harmonic decomposition to express the magnetic field components, we can investigate the energy distribution between different angular degrees $l$ to assess the field complexity. In Fig.~\ref{lmode} the magnetic field energy is shown as a function of $l$ for each ZDI map. It turns out that about 89-97\% of the total field energy is deposited in $l$=1-3 for all stars and observations. The dipole component, $l=1$, contains about 50-85\% of the total field energy. The smallest and largest values can be found for the same star, HN~Peg. In 12 out of the 13 observations of the four youngest stars, EK~Dra, HN~Peg, $\pi^1$~UMa and $\chi^1$~Ori, the energy is decreasing as $l$ increases from 1 to 3. The exception is the 2008.1 epoch of $\chi^1$~Ori where the energy of $l=2$ and $l=3$ is 6.4\% and 10.1\% respectively. This behaviour of decreasing energy with increasing $l$ is not seen for the two oldest stars, BE~Cet and $\kappa^1$~Cet where the energy of the octupole component is about twice as large as the energy of the quadrupole component.

The field topology can be expressed in terms of relative contributions of poloidal/toroidal components and axisymmetric/non-axisymmetric components. Here, axisymmetric is defined as all modes with $m < l/2$ \citep{Fares2009}. \citet{See2015} recently presented a study of 55 cool stars, including some of the stars in our study, where a correlation could be found between the toroidal and axisymmetric component of the magnetic field. $E_{\rm axis}/E_{\rm tot}$ seems to always be larger or approximately equal to $E_{\rm tor}/E_{\rm tot}$. Even though we have not used as strict definition of axisymmetry as was used in \citet{See2015} our results are consistent with that study as can be seen in Fig.~\ref{topo}.

The field is dominantly toroidal in half of our ZDI maps and close to 50\% in another three. This value varies by about 35\% for $\chi^1$~Ori where the field is dominantly toroidal in three observations and poloidal in one. It also varies by about 30\% for HN~Peg where $E_{\rm tor}/E_{\rm tot}$ is never above about 50\%. The fields are dominantly axisymmetric in 14 out of 16 magnetic maps. BE~Cet stands out since both the relative contributions of the toroidal component and the axisymmetric component are the smallest of all stars.

\section{Discussion}

In previous studies \citep[e.g.][]{Vidotto2014} it has been shown that the magnetic field strength decreases with increasing age, or rather, with increasing rotation period. In this study, constrained to a sample of young solar analogues, we see a significant decrease in the field energy and field strength when comparing EK~Dra, which is the youngest and most rapidly rotating star, to all the other stars. The increase in age is about 150~Myr between EK~Dra and HN~Peg, being the second youngest star, and the rotation period is increased by about 75\%. However, the difference between HN~Peg and the oldest star in this sample, $\kappa^1$~Cet, is about 400~Myr and the rotation period is doubled, but there is no decrease in the field strength comparing $\kappa^1$~Cet to HN~Peg. In fact, all stars older than EK~Dra show a similar field strength which varies as much for a single star with multiple epochs as it does between different stars. This sample is, on the other hand, quite small, and the significant difference between EK~Dra and the other stars could simply be because EK~Dra is a particularly active star. 

Interesting to note is that the same behavior is also seen for $\log L_{\rm X}$ and $\log R'_{\rm HK}$. There is a relatively large difference between EK~Dra and the older stars which all have similar or overlapping values. This is in line with the results from \citet{Schrijver1989}, who reported a power law relation between solar photospheric magnetic flux density and the local Ca\,{\sc ii} K emission. In the far-UV study by \citet{Fossati2015}, a similar trend was found for the C\,{\sc iv} 1548/C\,{\sc i} 1657 flux ratio of stars within the same age range as in our study. Their entire sample includes stars up to 7700~Myr with a gap in age between about 550~Myr and 4000~Myr where another significant decrease is seen. The C\,{\sc iv} 1548/C\,{\sc i} 1657 flux ratio for the entire sample follows a log-linear decrease with age. A log-linear decrease of $\langle B_{\rm r} \rangle$ with increasing age is also seen in the study by \citet{Vidotto2014}. It would therefore be interesting to extend our sample of solar analogues with older stars to see whether the same trend is seen for the mean magnetic field strength.

An interesting comparison can be made with HD~29615 which was included in our observational program ``Active Suns'' \citep{Hackman2015}. HD 29615 is believed to have solar-like parameters, $T_{\rm eff}=5866~K$, $M=0.97~M_{\odot}$ and $R=1.0~R_{\odot}$ \citep{McDonald2012,Allende-Prieto1999}, but a $P_{\rm rot}=2.32~d$ and age of about 30~Myr \citep{Messina2011,Zuckerman2011}. However, we did not include it in this study since we wanted a sample of well-studied, confirmed solar-analogues. HD 29615 lacks, for example, X-ray studies which is an important activity indicator. This star was analysed by \citet{Hackman2015} using the same ZDI code as in our study. The resulting magnetic maps show a similar mean magnetic field strength and energy as the 2007.1 epoch of EK~Dra, but a 1.5-3 times stronger $\langle B_{\rm r} \rangle$ compared to EK~Dra, while $\langle B_{\rm a} \rangle$ is 1.5-2 times weaker. The field is also dominantly poloidal, 63\%, in contrast to EK~Dra. This suggests a similar activity level as EK~Dra, but a different magnetic field configuration. \citet{Waite2015} also studied this star using a data set from 2009. Their results also show a similar mean magnetic field strength to EK~Dra and a dominantly poloidal field. In a recent study \citet{Folsom2016} found that stars with an age of about 120~Myr have a large scatter in the magnetic field strength and rotation period. Moreover, they did not see a trend where a shorter rotation period correlates with a stronger magnetic field. However, it should be noted that the range in effective temperature for these stars is about 4500 -- 5250~K and 0.75 -- 1.05~$M_\odot$ in mass. The authors also point out that this implies a range in convective zone depth which could explain the difference in magnetic field strength. 

The observed variation of the mean longitudinal field strength, \bz \ , hints that the magnetic field topologies are complex. As discussed at the end of section~\ref{multi}, the amplitude of \bz \ depends on magnetic field strength but also configuration. The range in absolute value of the maximum \bz \ strength, $|$\bz$|$, for each star with ascending age is 32.0--40.5~G, 7.5--16.0~G, 21~G, 4.6--7.5~G, 9.4~G and 6.2--8.2~G. This implies \bz \ is decreasing with increasing age with the exception of $\chi^1$~Ori. However, the reason for this is probably due to the lower overall field strength of $\chi^1$~Ori compared to the other stars. BE~Cet and $\kappa^1$~Cet, being the two oldest stars in this sample, have, on the other hand, comparable $\langle B \rangle$ strengths to the younger stars, except EK~Dra, but their \bz \ seems to be lower.  This implies that these two stars have a different field topology.

Another interesting feature seen in our ZDI results is the distribution of magnetic energy over different harmonic modes. The two oldest stars, BE~Cet and $\kappa^1$~Cet have an octupole component twice as large as the quadrupole component. The younger stars, with the exception of one of the images of $\chi^1$~Ori, all have a larger $l$=2 component compared to $l$=3. This behaviour should be confirmed by more observations of BE~Cet and $\kappa^1$~Ori, but also of other stars of similar age. It would also be interesting to observe stars older than 650~Myr to see if another trend arises and if so at what age. 

$\chi^1$~Ori displays polarity switches during the almost five years it was observed. The radial field on the pole changes polarity, but the meridional field also changes it's structure seemingly periodic. Since it was not observed during 2009, the magnetic cycle could be either 2, 6 or 8 years assuming periodicity. Follow up observations of this star are needed to confirm the magnetic cycle and determine its precise length. No polarity switch was seen for HN~Peg, which was observed during six years and constantly showed a pole with a positive radial field. It could, however, have a magnetic cycle that is longer than 12 years. $\kappa^1$~Cet was observed twice during a period of just under one year and the two observations of EK~Dra were separated by five years. $\kappa^1$~Cet shows the same polarity in both observations and so does EK~Dra, meaning that none of them can have a 2 year cycle. In other words, the stars in our sample do not all have the same magnetic cycle, if any, and it is not necessarily equal to the solar cycle either. This could mean the Sun used to have a different cycle length when it was younger. 

The global field of the Sun today is dominantly poloidal and it has been suggested in previous studies that the field topology changes with age, or rather, rotation rate and that slowly rotating stars tend to have a dominant poloidal component \citep[e.g.][]{Petit2008}. Rapidly rotating stars, on the other hand, do not seem to be coherently poloidal or toroidal \citep{Folsom2016}. In our study, 8 out of the 16 observations, show a dominantly toroidal field and for three observations the fraction is close to 50\%, as can be seen in Fig.~\ref{topo}. However, the youngest and oldest star in this study seem to have similar field configurations. The field topologies also vary significantly for the same star at different observation epochs meaning there are no definitive correlations between age and poloidal/toroidal component fractions for the age range studied here. The same conclusion can be drawn about the age dependence of the axisymmetric/non-axisymmetric components in this sample. 

The meridional field has the lowest local maximum strength in 10 of the maps and lowest mean field strength in all maps but one. Either this is a sign of a preferred magnetic field configuration for young solar-like stars or, more likely, it is again an indication that Stokes $V$ alone will often underestimate the strength of the meridional field compared to when Stokes $QU$ observations are included in ZDI inversions \citep{Rosen12,Rosen2015}. As discussed in section~\ref{zdi} the Stokes $V$ parameter is only sensitive to the line-of-sight component of the field. This implies that a meridional field vector will be almost perpendicular to the line-of-sight at all rotational phases. At the same time, the azimuthal field has the strongest local field strength in 9 out of 16 maps and strongest mean magnetic field in 13 out of 16 maps. This indicates that the azimuthal component is free of cross-talk, i.e. not misinterpreted as a radial or meridional component.

The other issues discussed in section~\ref{zdi} of only using Stokes $V$ to reconstruct the magnetic field could also influence the reliability of our results. The field strength could for instance be underestimated. This would of course influence results for individual stars, but probably the collective results would still be similar. The inclusion of linear polarisation could perhaps also result in  more complex magnetic field configurations. The question is, how much more complex. For Ap stars, including linear polarization does make the field more complex \citep[e.g.][]{Kochukhov10,Silvester2015}, but not as dramatic as for the cool RS~CVn star II~Peg \citep{Rosen2015}. It should, however, be noted that the magnetic field of II~Peg is more complex also in the Stokes $IV$ inversions compared to the magnetic fields derived in this study and to the Stokes $IV$ maps derived for Ap stars. This implies that even though the fields of the stars in our study would perhaps be more complex if linear polarisation could be included, the difference might not be as dramatic as in the case of II~Peg.

\section{Conclusions}
\label{con}

We have reconstructed the surface magnetic field topologies of six young solar analogue stars with ages from 100 to 650~Myr using 16 sets of high-resolution spectropolarimetric observations in total. Our analysis shows that the magnetic field strength decreases significantly with an increasing age from 100 to 250~Myr and an increase in rotation period by 75\%. However, this trend does not seem to continue as the stellar age and rotation period increase further. Instead, we see a similar variation for the same star at different observation epochs compared to the variation between stars of different age. 

When looking at the magnetic field topology of these stars, it seems that the octupole component is stronger than the quadrupole component for two stars older than 600~Myr. This is only seen in 1 out of the 13 ZDI maps of the younger stars. 

Our results suggest that the field topology of rapidly rotating young solar-mass stars can be either dominantly toroidal or poloidal. We also see that the ratio of axisymmetric field energy to total energy is roughly equal to or larger than the ratio of toroidal field energy to total energy. 

We observe two polarity switches for one of our targets explainable by a magnetic cycle period of either 2, 6 or 8 years. Some of the other stars do not seem to have the same period.  

The mean meridional field is the weakest in all but one epoch. On the other hand, the mean azimuthal field is the strongest in 13 out of 16 observations. This could either be a sign of a preferred magnetic field topology for young solar analogue stars, or it can be an indication that magnetic field reconstructions without linear polarisation will systematically underestimate the meridional field component but the azimuthal component is recovered correctly.

\begin{acknowledgements}
O.K. acknowledges financial support from the Knut and Alice Wallenberg Foundation, the Swedish Research Council, and the G\"{o}ran Gustafsson Foundation. J.L. acknowledges financial support from the Academy of Finland Centre of Excellence ReSoLVe (grant No. 272157).
\end{acknowledgements}

\bibliographystyle{aa}
\bibliography{astro_ref_v1}

\Online

\begin{figure*}
\centering
\includegraphics[width=\textwidth]{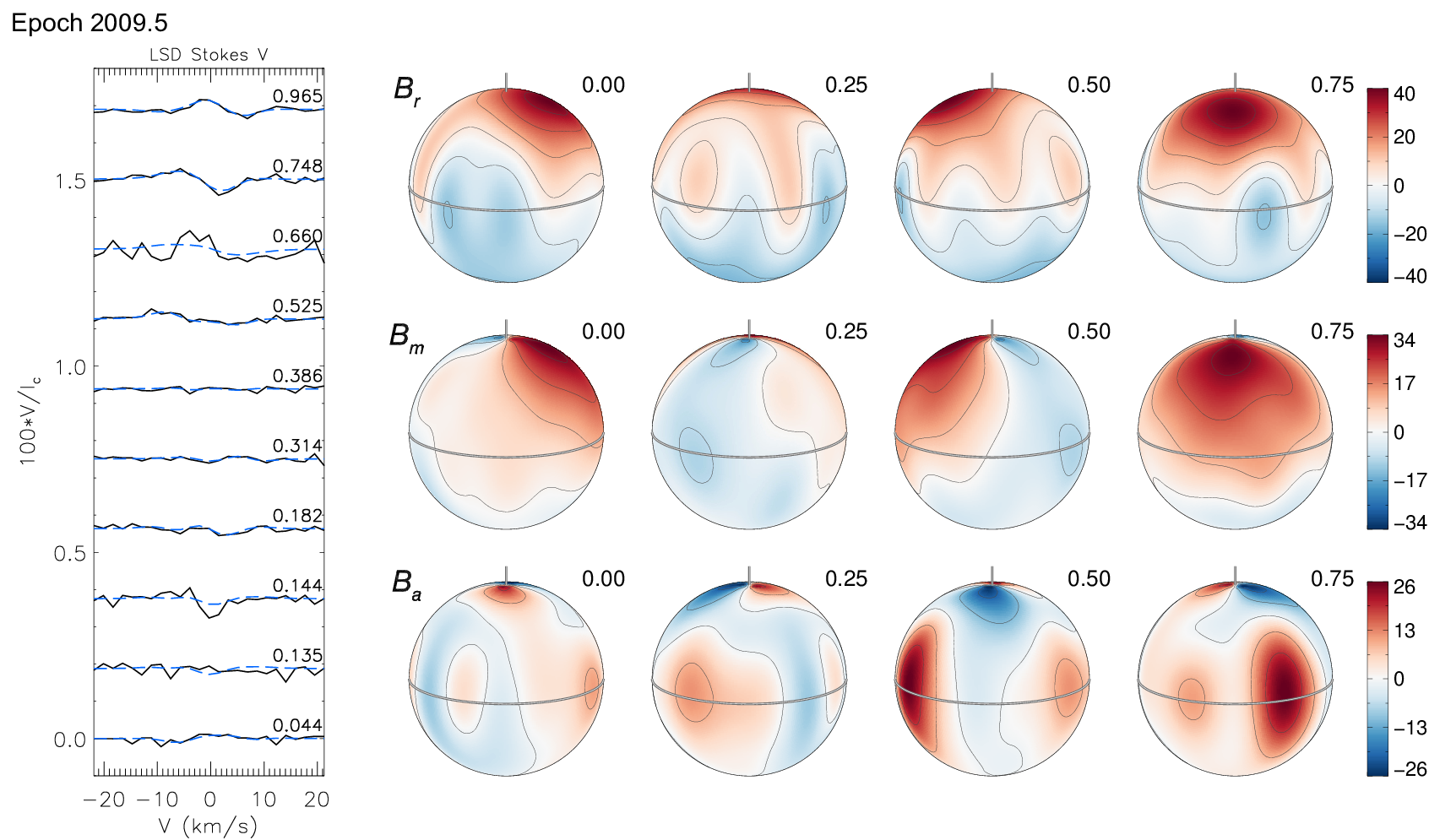}  
\includegraphics[width=\textwidth]{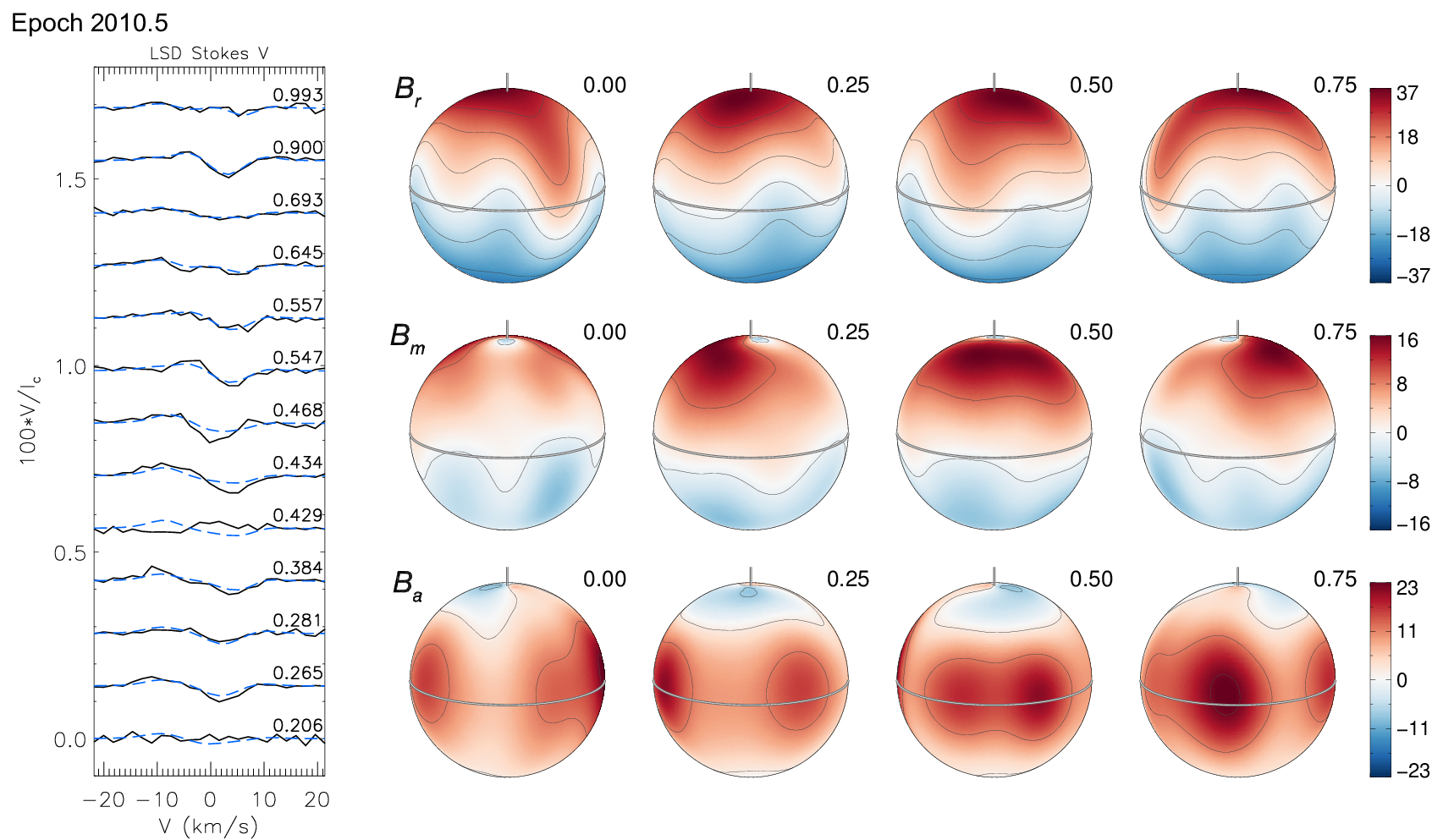} 
\caption{Same as Fig.~\ref{hn-map1} but here the line profiles and maps correspond to ZDI reconstructions for HN~Peg from the epoch 2009.5 epoch and the 2010.5 epoch.}
\end{figure*}

\begin{longtab}
\begin{longtable}{cccccc}
\caption{Log of spectropolarimetric observations of the six stars. The name of the star is found in the first column, the date of the observation in the second column, the heliocentric Julian date in the third column, the rotational phase in the fourth column, the associated mean uncertainty per velocity bin of the LSD profiles in the fifth column and the mean longitudinal magnetic field strength in the sixth column.}
\label{tab_obs}\\
\hline\hline

Star                  &  Date    & HJD        &   Rotational & $\sigma_{\rm LSD} \times 10^{-5}$ &{$\langle B_{\rm z} \rangle$}   \\
                         &  (UTC) & (2,400,000+) & phase   &                                                      &                     {(G)}\\
\hline
\endfirsthead
\caption{continued.}\\
\hline\hline

Star                  &  Date    & HJD        &   Rotational & $\sigma_{\rm LSD} \times 10^{-5}$ & {$\langle B_{\rm z} \rangle$}   \\
                         &  (UTC) & (2,400,000+) & phase   &                                                      &                    {(G)}\\
\hline
\endhead
\hline
\endfoot
EK~Dra & 2007-01-25 &   54126.7763 &    0.000 &  14.6 &   -32.0 $\pm$    9.0 \\
EK~Dra & 2007-01-26 &   54127.7412 &    0.371 &  13.4 &    10.8 $\pm$    8.1 \\
EK~Dra & 2007-01-27 &   54128.7197 &    0.747 &  12.4 &   -20.0 $\pm$    7.7 \\
EK~Dra & 2007-01-28 &   54129.7183 &    0.132 &  10.2 &     8.6 $\pm$    6.2 \\
EK~Dra & 2007-01-29 &   54130.7125 &    0.514 &  16.3 &    30.1 $\pm$    9.9 \\
EK~Dra & 2007-02-01 &   54133.7454 &    0.680 &  11.8 &     7.2 $\pm$    7.3 \\
EK~Dra & 2007-02-02 &   54134.7298 &    0.059 &  11.0 &    -9.6 $\pm$    6.8 \\
EK~Dra & 2007-02-03 &   54135.7418 &    0.448 &  10.3 &    25.9 $\pm$    6.2 \\
EK~Dra & 2007-02-18 &   54150.6318 &    0.175 &  13.9 &    -1.9 $\pm$    8.4 \\
EK~Dra & 2007-02-21 &   54153.6019 &    0.318 &  11.2 &    -9.9 $\pm$    6.9 \\
EK~Dra & 2007-02-22 &   54154.6318 &    0.714 &  11.5 &   -13.8 $\pm$    7.0 \\
\hline
EK~Dra & 2012-01-11 &   55938.7569 &    0.916 &   9.7 &     4.7 $\pm$    5.9 \\
EK~Dra & 2012-01-12 &   55939.7327 &    0.291 &  10.4 &   -19.6 $\pm$    6.3 \\
EK~Dra & 2012-01-13 &   55940.7212 &    0.671 &  12.0 &    40.5 $\pm$    7.4 \\
EK~Dra & 2012-01-14 &   55941.7209 &    0.056 &  12.0 &     7.2 $\pm$    7.3 \\
EK~Dra & 2012-01-15 &   55942.7082 &    0.435 &  12.6 &   -12.9 $\pm$    7.6 \\
EK~Dra & 2012-01-16 &   55943.6905 &    0.813 &  15.9 &    10.7 $\pm$    9.8 \\
EK~Dra & 2012-01-17 &   55944.6852 &    0.196 &  11.8 &    13.2 $\pm$    7.2 \\
EK~Dra & 2012-01-22 &   55949.7484 &    0.143 &  10.9 &    16.5 $\pm$    6.6 \\
EK~Dra & 2012-01-23 &   55950.7116 &    0.514 &  12.0 &    12.7 $\pm$    7.3 \\
EK~Dra & 2012-02-08 &   55966.5584 &    0.608 &  12.6 &     6.0 $\pm$    7.7 \\
\hline
HN~Peg & 2007-07-27 &   54309.5950 &    0.000 &   3.3 &    -1.1 $\pm$    1.1 \\
HN~Peg & 2007-07-28 &   54310.6025 &    0.219 &   4.5 &    -1.3 $\pm$    1.5 \\
HN~Peg & 2007-07-29 &   54311.5604 &    0.427 &   5.0 &    10.7 $\pm$    1.7 \\
HN~Peg & 2007-07-30 &   54312.5472 &    0.642 &   4.7 &     4.3 $\pm$    1.6 \\
HN~Peg & 2007-07-31 &   54313.5484 &    0.859 &   6.4 &     0.7 $\pm$    2.2 \\
HN~Peg & 2007-08-02 &   54315.5547 &    0.296 &   5.0 &     1.9 $\pm$    1.7 \\
HN~Peg & 2007-08-03 &   54316.5511 &    0.512 &   5.0 &     9.5 $\pm$    1.7 \\
HN~Peg & 2007-08-04 &   54317.5527 &    0.730 &   6.2 &     5.1 $\pm$    2.1 \\
HN~Peg & 2007-08-08 &   54321.5255 &    0.594 &   5.4 &    10.3 $\pm$    1.8 \\
HN~Peg & 2007-08-09 &   54322.5454 &    0.815 &   9.1 &     1.8 $\pm$    3.1 \\
HN~Peg & 2007-08-10 &   54323.5454 &    0.033 &   6.9 &     0.0 $\pm$    2.3 \\
HN~Peg & 2007-08-14 &   54327.5349 &    0.900 &   5.7 &     4.5 $\pm$    1.9 \\
HN~Peg & 2007-08-17 &   54330.5247 &    0.550 &   5.0 &    12.4 $\pm$    1.7 \\
HN~Peg & 2007-08-18 &   54331.4815 &    0.758 &   4.9 &     1.2 $\pm$    1.6 \\
\hline
HN~Peg & 2008-08-10 &   54689.5325 &    0.595 &   5.4 &    -3.2 $\pm$    1.8 \\
HN~Peg & 2008-08-12 &   54691.5126 &    0.026 &   4.9 &     4.0 $\pm$    1.6 \\
HN~Peg & 2008-08-17 &   54696.5413 &    0.119 &   7.5 &    -3.0 $\pm$    2.5 \\
HN~Peg & 2008-08-19 &   54698.6085 &    0.568 &   5.7 &     5.1 $\pm$    1.9 \\
HN~Peg & 2008-08-20 &   54699.5700 &    0.777 &   4.9 &     7.5 $\pm$    1.7 \\
HN~Peg & 2008-08-21 &   54700.4991 &    0.979 &   7.0 &     0.6 $\pm$    2.4 \\
HN~Peg & 2008-08-22 &   54701.4546 &    0.187 &   9.9 &    -4.1 $\pm$    3.3 \\
HN~Peg & 2008-08-23 &   54702.5291 &    0.420 &   4.2 &    -0.2 $\pm$    1.4 \\
HN~Peg & 2008-08-24 &   54703.5005 &    0.632 &   4.2 &     2.5 $\pm$    1.4 \\
HN~Peg & 2008-08-25 &   54704.5531 &    0.860 &   4.4 &     3.1 $\pm$    1.5 \\
\hline
HN~Peg & 2009-06-01 &   54984.6341 &    0.748 &   5.9 &     5.5 $\pm$    2.0 \\
HN~Peg & 2009-06-02 &   54985.6335 &    0.965 &   5.5 &    -1.4 $\pm$    1.8 \\
HN~Peg & 2009-06-03 &   54986.6306 &    0.182 &   5.3 &     4.7 $\pm$    1.8 \\
HN~Peg & 2009-06-12 &   54995.6157 &    0.135 &  17.3 &     7.8 $\pm$    5.8 \\
HN~Peg & 2009-06-19 &   55002.6326 &    0.660 &  14.7 &     7.8 $\pm$    4.9 \\
HN~Peg & 2009-06-22 &   55005.6399 &    0.314 &   6.0 &     3.2 $\pm$    2.0 \\
HN~Peg & 2009-06-23 &   55006.6102 &    0.525 &   6.7 &     5.8 $\pm$    2.2 \\
HN~Peg & 2009-06-27 &   55010.5685 &    0.386 &   6.2 &    -1.7 $\pm$    2.1 \\
HN~Peg & 2009-06-30 &   55013.5983 &    0.044 &   5.9 &    -4.0 $\pm$    2.0 \\
HN~Peg & 2009-07-05 &   55018.6589 &    0.144 &  11.1 &     3.3 $\pm$    3.8 \\
\hline
HN~Peg & 2010-06-21 &   55369.5907 &    0.434 &   6.1 &    13.8 $\pm$    2.0 \\
HN~Peg & 2010-07-04 &   55382.6157 &    0.265 &   4.8 &     6.9 $\pm$    1.6 \\
HN~Peg & 2010-07-06 &   55384.5849 &    0.693 &   4.8 &     5.3 $\pm$    1.6 \\
HN~Peg & 2010-07-07 &   55385.5330 &    0.900 &   4.8 &     3.6 $\pm$    1.6 \\
HN~Peg & 2010-07-10 &   55388.5557 &    0.557 &   5.6 &     9.7 $\pm$    1.9 \\
HN~Peg & 2010-07-12 &   55390.5612 &    0.993 &   7.3 &     2.7 $\pm$    2.4 \\
HN~Peg & 2010-07-13 &   55391.5437 &    0.206 &   8.3 &     2.1 $\pm$    2.8 \\
HN~Peg & 2010-07-14 &   55392.5666 &    0.429 &   5.6 &    -5.7 $\pm$    1.9 \\
HN~Peg & 2010-07-18 &   55396.4859 &    0.281 &   4.9 &     2.3 $\pm$    1.6 \\
HN~Peg & 2010-07-23 &   55401.5610 &    0.384 &   4.6 &     9.9 $\pm$    1.5 \\
HN~Peg & 2010-08-02 &   55411.5104 &    0.547 &   5.7 &     8.3 $\pm$    1.9 \\
HN~Peg & 2010-08-07 &   55416.5619 &    0.645 &   5.2 &     4.6 $\pm$    1.7 \\
HN~Peg & 2010-08-20 &   55429.5471 &    0.468 &   7.3 &    10.8 $\pm$    2.4 \\
\hline
HN~Peg & 2011-07-11 &   55754.5901 &    0.129 &   7.6 &    -2.3 $\pm$    2.5 \\
HN~Peg & 2011-07-21 &   55764.6140 &    0.308 &   5.3 &    -2.9 $\pm$    1.7 \\
HN~Peg & 2011-07-22 &   55765.5798 &    0.518 &   5.4 &     7.7 $\pm$    1.8 \\
HN~Peg & 2011-08-08 &   55782.5125 &    0.199 &   6.4 &    -5.6 $\pm$    2.1 \\
HN~Peg & 2011-08-10 &   55784.5154 &    0.635 &   6.2 &     4.0 $\pm$    2.1 \\
HN~Peg & 2011-08-11 &   55785.5713 &    0.864 &   5.7 &    16.0 $\pm$    1.9 \\
HN~Peg & 2011-08-15 &   55789.5906 &    0.738 &   6.3 &     3.1 $\pm$    2.1 \\
HN~Peg & 2011-08-16 &   55790.5309 &    0.943 &   5.9 &     7.1 $\pm$    1.9 \\
\hline
HN~Peg & 2013-09-09 &   56545.5301 &    0.073 &   4.9 &     2.5 $\pm$    1.6 \\
HN~Peg & 2013-09-09 &   56545.6705 &    0.103 &   4.8 &     3.7 $\pm$    1.6 \\
HN~Peg & 2013-09-10 &   56546.5378 &    0.292 &   5.7 &     1.5 $\pm$    1.9 \\
HN~Peg & 2013-09-10 &   56546.5769 &    0.300 &   5.9 &     3.1 $\pm$    2.0 \\
HN~Peg & 2013-09-10 &   56546.6711 &    0.321 &   5.1 &     3.7 $\pm$    1.7 \\
HN~Peg & 2013-09-11 &   56547.5290 &    0.507 &   5.7 &     9.0 $\pm$    1.9 \\
HN~Peg & 2013-09-11 &   56547.6715 &    0.538 &   4.7 &    10.7 $\pm$    1.5 \\
HN~Peg & 2013-09-12 &   56548.5379 &    0.727 &   3.9 &     5.0 $\pm$    1.3 \\
HN~Peg & 2013-09-12 &   56548.5787 &    0.736 &   3.6 &     2.3 $\pm$    1.2 \\
HN~Peg & 2013-09-12 &   56548.6717 &    0.756 &   3.3 &     2.3 $\pm$    1.1 \\
HN~Peg & 2013-09-14 &   56550.5609 &    0.166 &   5.6 &     7.1 $\pm$    1.8 \\
HN~Peg & 2013-09-14 &   56550.6727 &    0.191 &   6.9 &     7.5 $\pm$    2.3 \\
HN~Peg & 2013-09-15 &   56551.6320 &    0.399 &   5.9 &     1.9 $\pm$    2.0 \\
HN~Peg & 2013-09-15 &   56551.6920 &    0.412 &   6.8 &     1.5 $\pm$    2.3 \\

\hline
$\pi^1$~Uma & 2007-01-21 &   54122.4826 &    0.000 &  16.7 &   -21.0 $\pm$    6.8 \\
$\pi^1$~Uma & 2007-01-26 &   54127.5046 &    0.025 &   6.7 &   -11.4 $\pm$    2.7 \\
$\pi^1$~Uma & 2007-01-27 &   54128.5114 &    0.230 &   4.3 &    -4.2 $\pm$    1.8 \\
$\pi^1$~Uma & 2007-01-29 &   54130.5058 &    0.637 &   3.7 &     0.2 $\pm$    1.5 \\
$\pi^1$~Uma & 2007-02-01 &   54133.5353 &    0.256 &   5.0 &    -4.7 $\pm$    2.0 \\
$\pi^1$~Uma & 2007-02-02 &   54134.5044 &    0.453 &   5.5 &    -3.4 $\pm$    2.2 \\
$\pi^1$~Uma & 2007-02-03 &   54135.5271 &    0.662 &   5.1 &    10.5 $\pm$    2.1 \\
$\pi^1$~Uma & 2007-02-04 &   54136.4970 &    0.860 &   4.7 &     0.9 $\pm$    1.9 \\
$\pi^1$~Uma & 2007-02-06 &   54138.4632 &    0.261 &   8.8 &    -1.4 $\pm$    3.6 \\
$\pi^1$~Uma & 2007-02-07 &   54139.4554 &    0.464 &   9.6 &    -5.6 $\pm$    4.0 \\
$\pi^1$~Uma & 2007-02-08 &   54140.5101 &    0.679 &   5.8 &     6.1 $\pm$    2.4 \\
$\pi^1$~Uma & 2007-02-09 &   54141.5087 &    0.883 &   6.6 &    -5.2 $\pm$    2.7 \\

\hline
$\chi^1$~Ori & 2007-01-26 &   54127.4023 &    0.000 &   5.0 &    -2.2 $\pm$    1.7 \\
$\chi^1$~Ori & 2007-01-27 &   54128.3749 &    0.191 &   4.3 &    -4.6 $\pm$    1.4 \\
$\chi^1$~Ori & 2007-01-29 &   54130.3697 &    0.584 &   3.0 &     1.3 $\pm$    1.0 \\
$\chi^1$~Ori & 2007-02-01 &   54133.3883 &    0.178 &   3.6 &    -0.4 $\pm$    1.2 \\
$\chi^1$~Ori & 2007-02-02 &   54134.3617 &    0.370 &   9.0 &     1.8 $\pm$    3.1 \\
$\chi^1$~Ori & 2007-02-03 &   54135.3646 &    0.567 &   3.0 &     1.5 $\pm$    1.0 \\
$\chi^1$~Ori & 2007-02-04 &   54136.3580 &    0.763 &   3.3 &    -1.1 $\pm$    1.1 \\
$\chi^1$~Ori & 2007-02-06 &   54138.3566 &    0.156 &  10.6 &    -3.0 $\pm$    3.6 \\
$\chi^1$~Ori & 2007-02-08 &   54140.3746 &    0.554 &   3.6 &     2.0 $\pm$    1.2 \\
\hline
$\chi^1$~Ori & 2008-01-21 &   54487.3425 &    0.854 &   4.3 &    -2.1 $\pm$    1.4 \\
$\chi^1$~Ori & 2008-01-21 &   54487.3590 &    0.858 &   3.5 &     1.1 $\pm$    1.2 \\
$\chi^1$~Ori & 2008-01-22 &   54488.3602 &    0.055 &   5.7 &     3.0 $\pm$    1.9 \\
$\chi^1$~Ori & 2008-01-23 &   54489.3612 &    0.252 &   4.0 &     1.6 $\pm$    1.3 \\
$\chi^1$~Ori & 2008-01-24 &   54490.3819 &    0.453 &   4.6 &    -4.4 $\pm$    1.5 \\
$\chi^1$~Ori & 2008-01-25 &   54491.3651 &    0.646 &   3.8 &    -1.0 $\pm$    1.3 \\
$\chi^1$~Ori & 2008-01-26 &   54492.3645 &    0.843 &   4.2 &     3.1 $\pm$    1.4 \\
$\chi^1$~Ori & 2008-01-27 &   54493.3772 &    0.042 &   3.9 &     1.4 $\pm$    1.3 \\
$\chi^1$~Ori & 2008-01-28 &   54494.4366 &    0.251 &   4.2 &     1.9 $\pm$    1.4 \\
$\chi^1$~Ori & 2008-01-29 &   54495.3889 &    0.438 &   3.3 &    -7.5 $\pm$    1.1 \\
$\chi^1$~Ori & 2008-02-02 &   54499.3847 &    0.225 &   3.2 &     0.3 $\pm$    1.1 \\
$\chi^1$~Ori & 2008-02-04 &   54501.3678 &    0.615 &   3.2 &    -6.6 $\pm$    1.1 \\
$\chi^1$~Ori & 2008-02-05 &   54502.3709 &    0.813 &   3.6 &     1.1 $\pm$    1.2 \\
$\chi^1$~Ori & 2008-02-06 &   54503.4041 &    0.016 &   3.2 &     0.5 $\pm$    1.1 \\
$\chi^1$~Ori & 2008-02-09 &   54506.3765 &    0.601 &   3.7 &    -4.8 $\pm$    1.2 \\
$\chi^1$~Ori & 2008-02-10 &   54507.3208 &    0.787 &   2.9 &     1.0 $\pm$    1.0 \\
$\chi^1$~Ori & 2008-02-11 &   54508.3805 &    0.996 &   3.0 &     2.7 $\pm$    1.0 \\
$\chi^1$~Ori & 2008-02-12 &   54509.3842 &    0.193 &   3.4 &    -2.3 $\pm$    1.1 \\
$\chi^1$~Ori & 2008-02-13 &   54510.3801 &    0.389 &   3.6 &    -5.8 $\pm$    1.2 \\
$\chi^1$~Ori & 2008-02-14 &   54511.3971 &    0.590 &   3.3 &    -4.8 $\pm$    1.1 \\
$\chi^1$~Ori & 2008-02-15 &   54512.3791 &    0.783 &   3.3 &    -0.9 $\pm$    1.1 \\
\hline
$\chi^1$~Ori & 2010-09-19 &   55459.6636 &    0.256 &   2.8 &    -1.0 $\pm$    0.9 \\
$\chi^1$~Ori & 2010-09-26 &   55466.6600 &    0.633 &   3.7 &     1.3 $\pm$    1.2 \\
$\chi^1$~Ori & 2010-09-27 &   55467.6954 &    0.837 &   3.0 &     1.1 $\pm$    1.0 \\
$\chi^1$~Ori & 2010-09-28 &   55468.7158 &    0.038 &   3.4 &     0.6 $\pm$    1.1 \\
$\chi^1$~Ori & 2010-10-04 &   55474.5821 &    0.193 &   4.3 &     6.9 $\pm$    1.4 \\
$\chi^1$~Ori & 2010-10-13 &   55483.6796 &    0.984 &   2.7 &    -2.5 $\pm$    0.9 \\
$\chi^1$~Ori & 2010-10-20 &   55490.7126 &    0.368 &   5.8 &    -3.0 $\pm$    1.9 \\
\hline
$\chi^1$~Ori & 2011-10-11 &   55846.6525 &    0.435 &   6.1 &     4.3 $\pm$    2.0 \\
$\chi^1$~Ori & 2011-10-13 &   55848.6628 &    0.831 &   2.9 &    -3.3 $\pm$    1.0 \\
$\chi^1$~Ori & 2011-10-14 &   55849.7083 &    0.037 &   2.9 &    -1.0 $\pm$    1.0 \\
$\chi^1$~Ori & 2011-10-25 &   55860.6708 &    0.195 &   3.4 &     5.0 $\pm$    1.1 \\
$\chi^1$~Ori & 2011-11-08 &   55874.5771 &    0.932 &   3.7 &    -2.7 $\pm$    1.3 \\
$\chi^1$~Ori & 2011-11-16 &   55882.5564 &    0.503 &   3.5 &     4.1 $\pm$    1.2 \\
$\chi^1$~Ori & 2011-11-21 &   55887.5239 &    0.481 &  10.5 &     2.4 $\pm$    3.5 \\
$\chi^1$~Ori & 2011-11-23 &   55889.6361 &    0.896 &   4.0 &    -2.2 $\pm$    1.3 \\
$\chi^1$~Ori & 2011-11-26 &   55892.6407 &    0.488 &   4.4 &     6.4 $\pm$    1.5 \\
$\chi^1$~Ori & 2011-11-27 &   55893.6552 &    0.688 &   3.5 &     3.4 $\pm$    1.2 \\
$\chi^1$~Ori & 2011-11-28 &   55894.6328 &    0.880 &   3.1 &    -4.7 $\pm$    1.0 \\
$\chi^1$~Ori & 2011-11-29 &   55895.5903 &    0.069 &   3.3 &    -1.6 $\pm$    1.1 \\

\hline

BE~Cet & 2013-09-09 &   56545.5796 &    0.000 &   5.1 &    -0.5 $\pm$    1.4 \\
BE~Cet & 2013-09-09 &   56545.6296 &    0.007 &   5.2 &    -0.4 $\pm$    1.5 \\
BE~Cet & 2013-09-09 &   56545.7436 &    0.021 &   3.9 &    -2.4 $\pm$    1.1 \\
BE~Cet & 2013-09-10 &   56546.6225 &    0.136 &   4.1 &    -1.8 $\pm$    1.1 \\
BE~Cet & 2013-09-10 &   56546.7318 &    0.151 &   5.0 &    -1.6 $\pm$    1.4 \\
BE~Cet & 2013-09-11 &   56547.5761 &    0.261 &   5.9 &     1.5 $\pm$    1.7 \\
BE~Cet & 2013-09-11 &   56547.6220 &    0.267 &   5.1 &    -0.4 $\pm$    1.4 \\
BE~Cet & 2013-09-11 &   56547.7310 &    0.281 &   6.4 &     1.3 $\pm$    1.8 \\
BE~Cet & 2013-09-12 &   56548.6238 &    0.398 &   3.4 &     9.4 $\pm$    1.0 \\
BE~Cet & 2013-09-12 &   56548.7314 &    0.412 &   3.3 &     8.9 $\pm$    0.9 \\
BE~Cet & 2013-09-14 &   56550.6174 &    0.659 &   7.1 &     6.4 $\pm$    2.0 \\
BE~Cet & 2013-09-15 &   56551.7612 &    0.808 &   6.7 &     4.0 $\pm$    1.9 \\

\hline
$\kappa^1$~Cet & 2012-10-01 &   56202.5218 &    0.000 &   5.5 &     8.2 $\pm$    0.9 \\
$\kappa^1$~Cet & 2012-10-02 &   56203.5191 &    0.108 &   4.4 &     4.7 $\pm$    0.7 \\
$\kappa^1$~Cet & 2012-10-03 &   56204.5265 &    0.218 &   7.9 &     2.5 $\pm$    1.3 \\
$\kappa^1$~Cet & 2012-10-04 &   56205.5385 &    0.328 &   4.8 &     3.1 $\pm$    0.8 \\
$\kappa^1$~Cet & 2012-10-05 &   56206.5252 &    0.435 &   5.6 &     3.4 $\pm$    1.0 \\
$\kappa^1$~Cet & 2012-10-12 &   56213.5563 &    0.199 &   8.0 &     4.7 $\pm$    1.4 \\
$\kappa^1$~Cet & 2012-10-13 &   56214.4813 &    0.300 &   5.2 &     2.6 $\pm$    0.9 \\
$\kappa^1$~Cet & 2012-10-23 &   56224.5461 &    0.394 &   7.2 &     4.9 $\pm$    1.2 \\
$\kappa^1$~Cet & 2012-10-28 &   56229.5581 &    0.939 &   5.6 &     1.3 $\pm$    1.0 \\
$\kappa^1$~Cet & 2012-10-31 &   56232.5049 &    0.259 &   5.8 &     5.3 $\pm$    1.0 \\
$\kappa^1$~Cet & 2012-11-06 &   56238.5733 &    0.919 &   4.8 &     1.9 $\pm$    0.8 \\
$\kappa^1$~Cet & 2012-11-12 &   56244.4962 &    0.562 &   6.7 &    -0.9 $\pm$    1.2 \\
$\kappa^1$~Cet & 2012-11-14 &   56246.5362 &    0.784 &   4.9 &     3.3 $\pm$    0.8 \\
$\kappa^1$~Cet & 2012-11-22 &   56254.4716 &    0.647 &   7.6 &     0.3 $\pm$    1.3 \\
\hline
$\kappa^1$~Cet & 2013-09-09 &   56545.7018 &    0.302 &   4.9 &     1.2 $\pm$    0.8 \\
$\kappa^1$~Cet & 2013-09-09 &   56545.9281 &    0.327 &   5.1 &     0.9 $\pm$    0.9 \\
$\kappa^1$~Cet & 2013-09-10 &   56546.7004 &    0.411 &  11.0 &     0.1 $\pm$    1.9 \\
$\kappa^1$~Cet & 2013-09-10 &   56546.9248 &    0.435 &   3.5 &    -1.0 $\pm$    0.6 \\
$\kappa^1$~Cet & 2013-09-11 &   56547.7017 &    0.520 &   6.4 &     3.2 $\pm$    1.1 \\
$\kappa^1$~Cet & 2013-09-11 &   56547.9248 &    0.544 &   5.9 &     1.9 $\pm$    1.0 \\
$\kappa^1$~Cet & 2013-09-12 &   56548.7013 &    0.628 &   4.7 &     4.0 $\pm$    0.8 \\
$\kappa^1$~Cet & 2013-09-12 &   56548.9226 &    0.652 &   5.9 &     4.0 $\pm$    1.0 \\
$\kappa^1$~Cet & 2013-09-15 &   56551.7258 &    0.957 &   9.7 &    -4.7 $\pm$    1.7 \\
$\kappa^1$~Cet & 2013-09-15 &   56551.8868 &    0.974 &   6.8 &    -4.4 $\pm$    1.2 \\
$\kappa^1$~Cet & 2013-09-15 &   56551.8980 &    0.976 &   8.6 &    -6.4 $\pm$    1.5 \\

\hline
\end{longtable}
\end{longtab}

\end{document}